\documentclass[12pt]{article}
\usepackage{amsfonts,amsmath,amssymb}
\usepackage{graphicx}
\usepackage{epstopdf}

\newfont{\twelvemsb}{msbm10 scaled\magstep1}
\newfont{\eightmsb}{msbm8}
\newfam\msbfam
\textfont\msbfam=\twelvemsb \scriptfont\msbfam=\eightmsb
\catcode`\@=11
\def\Bbb{\ifmmode\let\next\Bbb@\else
\def\next{\errmessage{Use \string\Bbb\space only in math mode}}\fi\next}
\def\Bbb@#1{{\fam\msbfam{{#1}}}}

\newcommand{\be}{\begin{equation}}
\newcommand{\ee}{\end{equation}}
\newcommand{\ba}{\begin{eqnarray}}
\newcommand{\ea}{\end{eqnarray}}

\begin{document}

\sloppy
\renewcommand{\thefootnote}{\fnsymbol{footnote}}
\newpage
\setcounter{page}{1} \vspace{0.7cm}
\begin{flushright}
26/05/08
\end{flushright}
\vspace*{1cm}
\begin{center}
{\bf Large spin corrections in ${\cal N}=4$ SYM sl(2): still a linear integral equation}\\
\vspace{1.8cm} {\large Diego Bombardelli, Davide Fioravanti and
Marco Rossi\footnote{Permanent address of M.Rossi after May, 2nd:
Dipartimento di Fisica dell'Universit\`a della Calabria and INFN,
Gruppo collegato di Cosenza.}\footnote{E-mail:bombardelli@bo.infn.it, fioravanti@bo.infn.it, rossi@lapp.in2p3.fr}}\\
\vspace{.5cm} {\em Sezione INFN di Bologna, Dipartimento di Fisica, Universit\`a di Bologna, \\
Via Irnerio 46, Bologna, Italy} \\
\end{center}
\renewcommand{\thefootnote}{\arabic{footnote}}
\setcounter{footnote}{0}

\begin{abstract}
{\noindent Anomalous} dimension and higher conserved charges in the
$sl(2)$ sector of ${\cal N}=4$ SYM for generic spin $s$ and twist
$L$ are described by using a novel kind of non-linear integral
equation (NLIE). The latter can be derived under typical situations
of the SYM sectors, i.e. when the scattering need not depend on the
difference of the rapidities and these, in their turn, may also lie
on a bounded range. Here the non-linear (finite range) integral
terms, appearing in the NLIE and in the dimension formula, go to
zero as $s\rightarrow \infty$. Therefore they can be neglected at
least up to the $O(s^0)$ order, thus implying a linear integral
equation (LIE) and a linear dimension/charge formula respectively,
likewise the 'thermodynamic' (i.e. infinite spin) case. Importantly,
these non-linear terms go faster than any inverse logarithm power
$(\ln s)^{-n}$, $n>0$, thus extending the linearity validity.
\end{abstract}

\vspace{3cm}

\newpage

\section{Introduction}

The AdS/CFT duality \cite{MWGKP} conjectures the equivalence between
a string theory on the curved space-time
$\text{AdS}_5\times\text{S}^5$ in the strong coupling and a
conformal quantum field theory on the boundary of $\text {AdS}_5$ in
the weak coupling. In particular, type IIB superstring theory should
be dual to ${\cal N}=4$ Super Yang-Mills theory (SYM) in four
dimensions. In particular, it would relate energies of string states
to anomalous dimensions of local gauge invariant operators of the
quantum field theory. In this context, the discovery of
integrability in both free string theory and planar field theory was
a great achievement, both from the conceptual and the practical
(i.e. calculative) point of view, being the 't Hooft coupling,
$\lambda = 8 \pi^2 g^2$, the only non-running constant on stage (the
string tension is proportional to $g$). Actually, integrable models
appear as spin chain like Bethe equations, satisfied by 'rapidities'
which parametrise on the one side the quantum string states (and
their energies) and on the other side the corresponding composite
operators (and their anomalous dimensions) in SYM, respectively.
Actually, the initial result, which re-launched attention on
integrability in maximally SYM, identified the one-loop dilatation
operator of scalar gauge-invariant fields (of bare dimension $L$)
with a genuine $so(6)$ integrable hamiltonian of a spin chain (with
$L$ sites) \cite{MZ}. In the next turmoil, integrable structures
were hinted and found in all the sectors of ${\cal N}=4$ SYM and at
all loops (cf. for instance \cite {BKS}), bearing in mind the
convergence of the anomalous dimension (weak coupling) expansion.
Yet, all these integrable (Bethe, asymptotic) scattering equations
have the form of a deformation of the one-loop spin chain case, plus
an additional universal (string) scattering factor; the deformation
is such that the interaction range increases with the number of
loops. Therefore, starting from a certain loop order (generically
$L$ or higher), they are plagued by the 'wrapping' problem (cf. for
instance the third of \cite{BKS}), which was solved only in the
$SU(2)$ sector, where a mapping to the Hubbard model \cite{RSS} was
possible, however without incorporating the dressing factor. In a
parallel way, integrability in superstring theory was discovered at
classical level \cite{BPR} and then extended to semiclassical and
quantum level.

In this paper, we want contain ourselves within the (non-compact)
$sl(2)$ sector, viz. to the local composite operators
\begin{equation}
{\mbox {Tr}} ({\cal D}^s {\cal Z}^L)+.... \, , \label {sl2op}
\end{equation}
where ${\cal D}$ is the (symmetrised, traceless) covariant
derivative acting in all possible ways on the $L$ bosonic fields
${\cal Z}$. The spin of these operators is $s$ and $L$ is the
so-called 'twist'. Proper superpositions of operators (\ref{sl2op})
have definite anomalous dimension $\Delta$ depending on $L$, $s$ and
the `t Hooft coupling $\lambda = 8 \pi^2 g^2$:
\begin{equation}
\Delta = L+s+\gamma (g,s,L) \, , \label{Delta}
\end{equation}
where the anomalous part $\gamma$ is easily related to the
integrable chain energy, $E$ (not to be confused with the string
energy) via the well-known proportionality
\begin{equation}
\gamma (g,s,L)=g^2 E(g,s,L).
\end{equation}
Actually, at one loop the integrable problem is equivalent to the
analogue in planar one-loop QCD for various types of quasi-partonic
operators under specific circumstances \cite{LIP, BDM}
\footnote{Albeit QCD is in the whole not a conformal quantum field
theory, it still behaves like one at one loop and as far as the
anomalous dimensions are concerned. Even under these circumstances,
integrability is not complete since it requires additional
constraints as, for instance, aligned helicities of the partonic
degrees of freedom.}.

In the context of Bethe Ansatz like equations, a useful tool to
perform calculations is indeed the so-called non-linear integral
equation (NLIE) in its 'excited state' version \cite{FMQR}. The NLIE
allows to write exact expressions for the eigenvalues of the
observables for arbitrary values of the system length and of the
Bethe root number. Actually, this equation turns out to be more
efficient for numerical computations as well as for analytic
evaluations in some particular conditions, e.g. large number of
Bethe roots. Implementing this idea we previously found and
discussed finite size effects to the anomalous dimensions in the
$su(2)$ sector of ${\cal N}=4$ SYM \cite {FFGR1,FFGR2,FR}. In this
case, the NLIE and the exact expressions for the eigenvalues of the
charges have the same structure as in models studied in the past.
This is a consequence of two simple facts concerning this specific
case. First one, the scattering matrix between two magnons, which
appears in the r.h.s. of the Bethe equations, depends only on the
difference of their rapidities (Bethe roots), provided the so-called
dressing factor (cf. below) is neglected. Secondly, the Bethe roots
completely fill the real axis and only permit the presence of a
finite number of holes (or complex roots). However, at least one of
these properties fails when considering other sectors of ${\cal
N}=4$ SYM, or when the $S$-matrix is suitably equipped by a string
theory CDD factor \footnote{Although originally thought of as a
correction to the scattering coming from string effects, in the end
it revealed its effects already at four loops.}, the dressing term
\cite{BES}: this is indeed the situation in the $sl(2)$ sector. In
general, for dealing with this intricated structure of the Bethe
equations appearing in the whole ${\cal N}=4$ SYM, we proposed in
\cite{BFR} a path to a NLIE substantially different from the
original idea of \cite{FMQR}. The different strategy is to perform
the integrations just on the region (generally intervals of the real
axis) in which Bethe roots concentrate and, consequently, to avoid
the use of the Fourier transform in order to write the equation. It
follows that this new procedure is effective when the magnon
scattering matrix has a general dependence on the rapidities and the
Bethe roots are concentrated on intervals of the real axis or even
complex lines, i.e. in all the relevant cases of ${\cal N}=4$ SYM.
As a consequence, we obtain a simplification that we will explain
here in the $sl(2)$ case: since nonlinear integral terms enjoy an
integration just on the region where the $s$ Bethe roots actually
lie, they become depressed more than any inverse power $(\ln
s)^{-n}$, $n=1,2,\dots$, for any fixed value of the twist $L$.
Therefore, stated in advance the logarithmic scaling
\begin{equation}
\gamma (g,s,L)=f(g)\ln s + O(s^0) \, ,
\end{equation}
after the very important BES' paper \cite{BES} on $f(g)$ (cf. also
e.g. \cite{KM} for preceding literature), we are left with a Linear
Integral Equation (LIE) which allows us to compute the sub-leading
corrections, $O(s^0)$, of the conformal dimensions \footnote{Then,
we could also consider the limit $L= j \ln s\rightarrow \infty$ for
fixed $j$, which is indeed the relevant scaling of this theory
\cite{BGK} and entails an improvement of the previous formula into
$\gamma (g,s,L)=f(g,j)\ln s + O((\ln s)^{-\infty})$ \cite{FRS}; but
this would be the subject of some future publications.} and, of the
other charges, both the leading and the subleading terms. In this
respect, our approach is clearly different from that of \cite{FRS},
which uses the full real axis NLIE presentation by \cite{FMQR},
because this needs to take into account and to evaluate the
non-linear integrals (on the whole real axis) as well.

Very interestingly, our LIE does not differ from the BES one
\cite{BES}, but for the inhomogeneous part, which consists in an
integral on the one loop root density and a hole depending term
(apart from a known function) \footnote{This structure encourages us
to proceed further in the direction of the fixed $j$ expansion and
suggests the preservation of itself: but this way may only be the
topic of future publications.}. As the equation is non-perturbative
(i.e. for any $g$) and linear and it drives the cross-over between
weak (small $g$) to strong (large $g$) coupling in a rather
intelligible way. Nevertheless, for evaluating and checking dominant
string effects (like for instance the dressing phase) the
aforementioned wrapping effects ought to be negligible or known.

Eventually, these facts have furnished us the stimulus to
investigate the next-to-leading-order (nlo) term -- although coming
from an asymptotic Bethe Ansatz -- in that the leading order $f(g)$
has been conjectured to be independent of $L$ or {\it universal}
\cite{BES}, after the one loop proof by \cite{BGK}.

In this paper we will give an explicit application of this new type
of (N)LIE to the $(L,s)$-vacua of the $sl(2)$ sector with the
following plan. In Section 2, we will outline the formalism suitable
for writing the NLIE on an interval in the most general case. In
Section 3 we will apply this technique to the case of the spin
$-1/2$ XXX chain, which describes the one loop $sl(2)$ sector of
${\cal N}=4$ SYM and in Section 4 we will discuss the many loop case
and end up with our main object, the linear integral equation. By
means of the latter, we will compute up to three loops in the 't
Hooft coupling the leading and sub-leading corrections of the
eigenvalues of the conserved charges in the large $s$ limit, and
check our results vs. the anomalous dimensions in \cite{FRS} (for
what concerns the other charges, our findings are new at the best of
our knowledge).

\section{A new approach: the NLIE on the interval(s)}
\setcounter{equation}{0}

In almost all the cases considered up to now, the NLIE was written
for counting functions defined as \be Z(u)=\Phi (u)-\sum _{k=1}^{s}
\phi (u-u_k) \, , \ee and when the Bethe roots distribute on the
real axis, allowing the presence of only a finite number of holes
and possibly complex roots. Even if this case is relevant for the
study of the fundamental state and the first excitations of many
models, it does not cover many of the Bethe Ansatz systems proposed
in the context of ${\cal N}=4$ SYM.

For this reason we want to write the NLIE (and the expression for
the eigenvalues of the observables in terms of its solution) for the
more general case in which the counting function is defined as \be
Z(u)=\Phi (u)-\sum _{k=1}^{s} \phi (u,u_k) \, , \label {Z2} \ee
(i.e. the function $\phi (x,y) $ does not depend only on the
difference $x-y$: this happens, for instance, when the dressing
factor is present). We suppose also that the $s$ Bethe roots $\{ u_k
\}_{k=1,...,s}$ are concentrated in an interval $[A,B]$ of the real
axis \footnote {The case in which the Bethe roots are concentrated
on a finite number of intervals on the real axis follows
straightforwardly from the results of this Section. Moreover, even
the case when the roots lie on complex lines can be treated as
follows.} and that a finite number of holes is present. We call
$u^{(i)}_h$ the holes (in number $H_i$) lying inside the interval
and $u^{(o)}_h$ the holes (in number $H_o$) lying outside the
interval. This particular distribution of roots is peculiar, for
instance, of states in the $sl(2)$ sector of ${\cal N}=4$ SYM.

On this state we consider a sum over the Bethe roots $\{ u_k
\}_{k=1,...,s}$ of a function (observable) $O(u)$ analytic in a
strip around the real axis. Without loss of generality, we can put
ourselves in the case in which on both roots and holes the condition
$e^{iZ(u_k)}=e^{iZ(u_h^{(i)})}=-1$ holds. In this case, this sum can
be written \cite {FMQR} as
\begin{eqnarray}
&& 2\pi i \sum _{k=1}^{s}O(u _k)= \lim _{\epsilon \rightarrow 0^+} \left [ \int _{A}^{B}du  O(u-i\epsilon)\frac {e^{iZ(u-i\epsilon)}iZ^{'}(u-i\epsilon)}{1+e^{iZ(u-i\epsilon)}}+ \right. \label {equ2} \\
&+& \left. \int _{B}^{A}du  O(u+i\epsilon)\frac
{e^{iZ(u+i\epsilon)}iZ^{'}(u+i\epsilon)}{1+e^{iZ(u+i\epsilon)}}
\right ] -2\pi i \sum _{h=1}^{H_i} O(u_h^{(i)}) \, . \nonumber
\end{eqnarray}
Supposing $Z^{'}(u)<0$ and supposing that the values $Z(A)$ and
$Z(B)$ are known, we can rearrange this expression as follows,
\begin{eqnarray}
\sum _{k=1}^{s}O(u _k)&=&-\int _{A}^{B}\frac {dv}{2\pi}
O(v)Z^{'}(v)+\int _{A}^{B}\frac {dv}{\pi} O(v)\frac {d}{dv}{\mbox
{Im}}\ln \left
[1+e^{iZ(v-i0)}\right] - \nonumber \\
&-&\sum _{h=1}^{H_i} O(u_h^{(i)}) =  - \frac {1}{2\pi}\left [O(B)Z(B)-O(A)Z(A) \right] +\label {prop2} \\
&+&\frac {1}{\pi} \left \{ O(B) {\mbox {Im}}\ln \left
[1+e^{iZ(B)}\right]- O(A)
{\mbox {Im}}\ln \left [1+e^{iZ(A)}\right] \right \} + \nonumber \\
&+& \int _{A}^{B}\frac {dv}{2\pi} O^{'}(v) Z(v)- 2\int _{A}^{B}\frac
{dv}{2\pi} O^{'}(v){\mbox {Im}}\ln \left [1+e^{iZ(v-i0)}\right] -
\sum _{h=1}^{H_i} O(u_h^{(i)})\, . \nonumber
\end{eqnarray}
In brief, what we are doing is to evaluate a sum on the Bethe roots
by integrating just on the interval containing them. Therefore, this
method is alternative and complementary to the idea proposed in the
first of \cite {FMQR} which consists in first integrating on all the
real axis and then subtracting the contributions coming from the
real holes. The convenience of this new method is that the
non-linear terms present in (\ref {prop2}) are strongly suppressed
in the limit of large number of Bethe roots. We will come back on
this point in the next sections, when we shall apply this technique
to the $sl(2)$ sector of ${\cal N}=4$ SYM.

We now use (\ref {prop2}) in the sum over the Bethe roots appearing
in the definition (\ref{Z2}) and obtain the following equation
\begin{eqnarray}
Z(u)&=&\Phi (u) + \int _{A}^{B}\frac {dv}{2\pi} \, \phi (u,v)\, \frac {d}{dv}  Z(v)+\sum _{h=1}^{H_i}\phi (u,u_h^{(i)})-\nonumber \\
&-&2\int _{A}^{B}\frac {dv}{2\pi} \,  \phi (u,v)\, \frac {d}{dv}
{\mbox
{Im}}\ln \left [1+e^{iZ(v-i0)}\right] = \nonumber \\
&=&f(u) - \int _{A}^{B}\frac {dv}{2\pi} \, \frac {d}{dv} \phi (u,v)\, Z(v)+\nonumber \\
&+&2\int _{A}^{B}\frac {dv}{2\pi} \, \frac {d}{dv} \phi (u,v)\,
{\mbox {Im}}\ln \left [1+e^{iZ(v-i0)}\right] \, , \label{2eq1}
\end{eqnarray}
where
\begin{eqnarray}
f(u)&=&\Phi (u)+\frac {1}{2\pi}\left [ \phi (u,B)Z(B)-\phi (u,A)Z(A) \right] +\sum _{h=1}^{H_i}\phi (u,u_h^{(i)}) -\nonumber \\
&-&\frac {1}{\pi} \left \{ \phi (u,B) {\mbox {Im}}\ln \left
[1+e^{iZ(B)}\right]- \phi (u,A) {\mbox {Im}}\ln \left
[1+e^{iZ(A)}\right] \right \}  \, .
\end{eqnarray}
We can now write a NLIE for the counting function by inserting in an
iterative way (\ref {2eq1}) for $Z$ in the right hand side of the
same equation. Using the notation
\begin{equation}
(\varphi \star f) (u) = \int _{A}^{B} dv \, \varphi (u,v)f(v) \, ,
\end{equation}
eventually we gain the NLIE in the final form
\begin{equation}
Z(u)= F (u) + 2 (G \star L) (u) \, , \label {nlin}
\end{equation}
where
\begin{equation}
F (u)= f(u)+\sum _{k=1}^{\infty} (-1)^k ((\varphi ^{\star k}) \star
f )(u) \, , \quad G(u,v)=  \varphi (u,v)+ \sum _{k=2}^{\infty}
(-1)^{k-1} \, (\varphi ^{\star k} )(u,v) \, . \label {FGdef}
\end{equation}
We used the simplified notations
\begin{equation}
L(u)={\mbox {Im}}\ln \left [1+e^{iZ(u-i0)}\right] \, , \quad \varphi
(u,v) = \frac {1}{2\pi} \frac {d}{dv} \phi (u,v) \label{L} \, .
\end{equation}
More explicitly,
\begin{eqnarray}
F(u)&=& f(u)+\sum _{k=1}^{\infty}(-1)^k \int _{A}^{B}dv_1 \, \varphi (u,v_1)  \int _{A}^{B}dv_2 \, \varphi (v_1,v_2) \ldots \nonumber \\
&\ldots &
\int _{A}^{B}dv_k \, \varphi (v_{k-1},v_k) f(v_k) \, , \label {Fdef} \\
G(u,v)&=&\varphi (u,v)+\sum _{k=1}^{\infty} (-1)^k \int _{A}^{B}dv_0 \, \varphi (u,v_0)\int _{A}^{B}dv_1\, \varphi (v_0,v_1) \ldots \nonumber \\
&\ldots & \int _{A}^{B}dv_{k-1} \, \varphi (v_{k-2}, v_{k-1}) \,
\varphi (v_{k-1}, v) \, . \label {Gdef}
\end{eqnarray}
These expressions are quite formal and very difficult to handle.
More easily, both the {\it forcing term} $F(u)$ and the {\it kernel}
$G(u,v)$ may be found by solving respectively the linear integral equation, which easily follow
from (\ref {Fdef}, \ref {Gdef}),
\begin{eqnarray}
F(u)&=&f(u)-\int _{A}^{B}dv \varphi (u,v) F(v) \, ,\label {Feq}\\
G(u,v)&=&\varphi (u,v)-\int _{A}^{B}dw \varphi (u,w) G(w,v) \, , \label {Geq} \\
\end{eqnarray}
and are linked the one to the other via
\begin{equation}
F(u)=f(u)-\int _{A}^{B}dv G(u,v) f(v) \, . \label {FGeq}
\end{equation}

Eventually, inserting (\ref {nlin}) in (\ref {prop2}) we obtain an
expression for the eigenvalues of an observable as
\begin{eqnarray}
\sum _{k=1}^s O(u_k)&=&
- \frac {1}{2\pi}\left [O(B)Z(B)-O(A)Z(A) \right] +\nonumber \\
&+&\frac {1}{\pi} \left \{ O(B) {\mbox {Im}}\ln \left
[1+e^{iZ(B)}\right]- O(A)
{\mbox {Im}}\ln \left [1+e^{iZ(A)}\right] \right \} + \nonumber \\
&+& \int _{A}^{B}\frac {dv}{2\pi}   O^{'}(v) F(v)-\sum _{h=1}^{H_i}O(u_h^{(i)})+\label {Oexp} \\
&+&2\int _{A}^{B}\frac {dv}{2\pi}  O^{'}(v)\int _{A}^{B} dw
[G(v,w)-\delta (v-w)]{\mbox {Im}}\ln \left [1+e^{iZ(w-i0)}\right] \,
. \nonumber
\end{eqnarray}
We remark that all the already known NLIEs can be reproduced in this
way, without Fourier transforming. In this sense the method sketched
in this section is more general. It seems natural to use formul{\ae}
(\ref {nlin}, \ref {FGdef}) and (\ref {Oexp}) in order to write,
respectively, the NLIE and the eigenvalues of the observables on
states appearing in models relevant for ${\cal N}=4$ SYM. In the
next two sections we will apply this new method to the widely
studied \cite{BKS,ES,BES} $sl(2)$ sector of the theory.

\section{An example: the $XXX_{-\frac {1}{2}}$ spin chain}
\setcounter{equation}{0}

We want to apply the techniques developed in the previous Section to
the $XXX_{-\frac {1}{2}}$ spin chain, as the latter clearly gives a
representation of the $sl(2)$ sector at one loop, as stated in the
'Introduction'.

The spectrum of this non-compact spin chain may be described by the
Bethe equations,
\begin{equation}
\left(\frac{u_k-\frac{i}2}{u_k+\frac{i}2}\right)^L=
\mathop{\prod^s_{j=1}}_{j\neq k} \frac{u_k-u_j+i}{u_k-u_j-i} \, ,
\label {xxxbethe}
\end{equation}
where we have indicated with $L$ the length of the chain and $s$ the
number of Bethe roots. In this case the $s$ Bethe roots concentrate
in an the interval of the real axis symmetric with respect to zero.
$L$ holes are present: two holes $u_h^{(o)}$, $h=1,2$, lie outside,
$H_i=L-2$ holes $u_h^{(i)}$, $h=1,\ldots ,L-2$, lie inside this
interval. Let us define the counting function (for reasons that will
be clear in the following, we will put an index $0$ to all the
functions, e.g. $Z$, $F$, $G$, related to the $XXX_{-\frac {1}{2}}$
spin chain) as
\begin{equation}
Z_0(u)=iL \ln \left (\frac {\frac {i}{2}-u} {\frac {i}{2}+u} \right
)-i \sum _{j=1}^s \ln \left (\frac {i+u-u_j}{i-u+u_j} \right ) \, .
\end{equation}
From the Bethe equations (\ref {xxxbethe}), we have the conditions
\begin{equation}
iL \ln \left (\frac {u_k-\frac {i}{2}} {u_k+\frac {i}{2}} \right )-i
\mathop{\sum _{j=1}^s}_{j\neq k} \ln \left (\frac
{u_k-u_j+i}{u_k-u_j-i} \right )=2\pi n_k \, , \quad n_k \in {\mathbb
Z} \, .
\end{equation}
Using the property
\begin{equation}
 i\ln \left (\frac {x-{i}\xi} {x+{i}\xi} \right )-i\ln \left (\frac {{i}\xi-x} {{i}\xi+x} \right )=\pi \, , \quad \xi > 0 \, ,
\end{equation}
we have that, on both Bethe roots $u_k$ and holes $u_h$,
\begin{equation}
Z_0(u_k)=\pi(2n_k-L-s+1) \, , \quad Z_0(u_h)=\pi(2n_h-L-s+1) \, .
\end{equation}
We choose $L+s$ even, in such a way that
$e^{iZ(u_k)}=e^{iZ(u_h^{(i)})}=e^{iZ(u_h^{(o)})}=-1$, as in the
previous Section\footnote {The number of derivatives $s$ has to be even and in any case, there is no loss of generality, in that, if $L+s$ is odd, the only modification in all the formul{\ae} is the replacement
of the logarithmic indicator $L_0(u)$ by ${\mbox
{Im}}\ln \left [1-e^{iZ_0(u-i0)}\right]$.}. If we now go from the
smallest to the biggest root of the interval (passing also $L-2$
internal holes) the (decreasing) counting function $Z_0$ varies of
$-2\pi (s+L-3)$. Since $Z_0$ is an odd function, this means that
\begin{equation}
Z_0(u_{s})=-\pi (s+L-3) \, , \quad Z_0(u^{(o)}_1)=-\pi (s+L-1) \, ,
\end{equation}
where $u_{s}$ is the biggest (most positive) root of the interval
and $u^{(o)}_1$ is the positive hole outside the interval. We can
choose the separator $b_0$, which defines the interval of
integration in our formul{\ae}, i.e. $A=-b_0 \, , \quad B=b_0$, such
that
\begin{equation}
 Z_0(-b_0)=-Z_0(b_0)=\pi (s+L-2) \, . \label {Zcond}
\end{equation}
With this position, the relevant functions defined in the previous
section take the form
\begin{eqnarray}
\Phi _0(u)&=&-2L \arctan 2u \, , \ \phi _0(u,v)=2 \arctan (u-v) \, , \ \varphi _0(u,v)= -\frac {1}{\pi} \frac {1}{1+(u-v)^2} \, , \nonumber \\
\quad f_0(u)&=&-2L\arctan 2u - (s+L-2)[\arctan (u-b_0)+\arctan (u+b_0) ] + \label {sl2funct} \\
&+& 2 \sum _{h=1}^{L-2}\arctan (u-u_h^{(i)})\, . \nonumber
\end{eqnarray}
Since $s+L$ is even, we have that ${\mbox {Im}}\ln \left
[1+e^{iZ_0(\pm b_0)}\right] =0$. In addition, one easily shows that
in this case the functions $F _0$ and $G_0$ enjoy the parity
properties
\begin{equation}
F_0(u)=-F_0(-u) \, , \quad G_0(u,v)=-G_0(-u,v)=G_0(-u,-v) \, .
\label {FGparity}
\end{equation}
They may be determined by solving the linear equations
\begin{eqnarray}
F_0(u)&=& -2L\arctan 2u - (s+L-2)[\arctan (u-b_0)+\arctan (u+b_0) ]  + \nonumber \\
&+& 2 \sum _{h=1}^{L-2}\arctan (u-u_h^{(i)}) + \int _{-b_0} ^{b_0}
\frac {dv}{\pi} \frac {1}{1+(u-v)^2} F_0(v)  \, , \label {F1loop}
\end{eqnarray}
\begin{equation}
G_0(u,v)=-\frac {1}{\pi} \frac {1}{1+(u-v)^2} + \int _{-b_0}^{b_0}
\frac {dw}{\pi} \frac {1}{1+(u-w)^2} G_0(w,v)  \label{G1loop} \, .
\end{equation}
At the leading $s\rightarrow \infty$ order these equations become
simpler and therefore we will start with this case.

\subsection{Determination of $F_0$ when $s\rightarrow +\infty$}

Let us consider equation (\ref {F1loop}) for the function $F_0(u)$.
It is convenient to rescale the variable $u=\bar u s$ and the
extreme of the interval $b_0=\bar b _0s$, in view of the limit $s
\rightarrow +\infty$:
\begin{eqnarray}
F_0(\bar u s)&=& -2L\arctan 2\bar u s- (s+L-2)[\arctan (\bar us -\bar b_0 s)+\arctan (\bar u s+\bar b_0 s) ]  + \nonumber \\
&+& 2 \sum _{h=1}^{L-2}\arctan (\bar u s -u_h^{(i)}) + \int _{-\bar
b_0} ^{\bar b_0} d\bar v \frac {s}{\pi} \frac {1}{1+s^2 (\bar u
-\bar v)^2} F_0(\bar v s)  \, . \label {F1loop2}
\end{eqnarray}
Using the asymptotic expansions ($O(s^{-n})$ means ``terms of order
$s^{-n}$'')
\begin{eqnarray}
\arctan \bar u s &=& \frac {\pi}{2} {\mbox {sgn}} \bar u - \frac {1}{\bar u s} + O(s^{-3})  \label {functexp} \, \\
\frac {s}{\pi} \frac {1}{1+s^2 (\bar u -\bar v)^2}&=&\delta (\bar u
-\bar v) -\frac {1}{s\pi} \frac {d}{d\bar u} \frac {1}{\bar u -\bar
v} +O(s^{-3}) \, ,
\end{eqnarray}
{\it inside the integral}\footnote{This is indeed a subtle exchange of two limits.} and the fact that
\begin{equation}
\lim _{s\rightarrow +\infty} u_h^{(i)}=0 \, ,
\end{equation}
we obtain {\it in this approximation}
\begin{eqnarray}
F_0(\bar u s)&=& -2\pi {\mbox {sgn}}(\bar u) +\frac {4-L}{\bar u s}+\frac {s+L-2}{s}\frac {2\bar u}{\bar u^2 -\bar b^2_0} + F_0(\bar u s) + \nonumber \\
&+& \int _{-\bar b_0} ^{\bar b_0} d\bar v \frac {1}{s \pi} \left (
\frac {d}{d\bar v} P\frac {1}{\bar u -\bar v} \right )  F_0(\bar v s)   \, .
\end{eqnarray}
Integrating by part gives
\begin{eqnarray}
0&=& -2\pi {\mbox {sgn}}(\bar u)+\frac {4-L}{\bar u s} +\frac {s+L-2}{s}\frac {2\bar u}{\bar u^2 -\bar b^2_0} - \frac {F_0(-\bar b _0s)}{s\pi }\left (\frac {1}{\bar u +\bar b_0}+\frac {1}{\bar u -\bar b_0}\right ) - \nonumber \\
&-& \int _{-\bar b_0} ^{\bar b_0} d\bar v \frac {1}{s \pi} \left (
P\frac {1}{\bar u -\bar v} \right ) \frac {d}{d\bar v} F_0(\bar v s) \, . \label {eqint}
\end{eqnarray}
Now, condition (\ref{Zcond}) at large $s$ imposes on the function
$F_0$ the constraint
\begin{equation}
F_0(\bar b _0s)=-F_0(-\bar b _0s)=-\pi (s+L-2) + o(s^{0})\, , \label
{Fcond}
\end{equation}
where $o(s^0)$ means ``terms of order smaller than $s^0$''.
Therefore, equation (\ref{eqint}) simplifies into
\begin{equation}
0=-2\pi {\mbox {sgn}}(\bar u) +\frac {4-L}{\bar u s}- \int _{-\bar
b_0} ^{\bar b_0} d\bar v \frac {1}{s \pi} \left ( P\frac {1}{\bar u
-\bar v} \right ) \frac {d}{d\bar v} F_0(\bar v s) \, .
\end{equation}
This equation can be solved by finite Hilbert transform techniques.
Its solution reads
\begin{eqnarray}
-\frac {1}{2\pi s}\frac {d}{d\bar u} F_0(\bar u s)&=& \frac {1}{\pi} \ln \left ( \frac {\bar b_0 + {\sqrt {\bar b^2_0 -\bar u^2}}}{\bar u} \right )^2  -\frac {2-\frac {L}{2}}{s} \delta (\bar u) \Rightarrow  \label {dersol} \\
F_0(\bar u s) &=& -4\bar b_0 s \arcsin \frac {\bar u}{\bar b_0}  -2 \bar u s \ln \left ( \frac {\bar b_0 + {\sqrt {\bar b ^2_0 -\bar u^2}}}{\bar u} \right )^2  +\pi \left (2-\frac {L}{2}\right ){\mbox {sgn}}\bar u \nonumber \, . 
\end{eqnarray}
Remember, however, that also $\bar b_0$ depends on $s$ through (\ref
{Zcond}), which at large $s$ implies (\ref{Fcond}) - a
normalization condition for (\ref{dersol}) in this approximation
\begin{equation}
\int _{-\bar b_0}^{\bar b_0} d\bar u  \left [ \frac {1}{\pi} \ln
\left ( \frac {\bar b_0 + {\sqrt {\bar b^2_0 -\bar u^2}}}{\bar u}
\right )^2  -\frac { 2-\frac {L}{2}}{s} \delta (\bar u) \right ] =
1+ \frac {L-2}{s} \, .
\end{equation}
From this equation we deduce, after integration,
\begin{equation}
2\bar b _0-\frac {2-\frac {L}{2}}{s}=1+ \frac {L-2}{s}  \,
, \quad \Rightarrow  \bar b _0= \frac {1}{2}\left (1+\frac
{{L}}{2s}\right )   \, . \label {bexp}
\end{equation}
Inserting this expansion in (\ref {dersol}), we obtain the following
behaviour of $F_0$ and its derivative when $s\rightarrow \infty$:
\begin{eqnarray}
-\frac {1}{2\pi s}\frac {d}{d\bar u} F_0(\bar u s)&=& \frac {1}{\pi} \ln \left ( \frac {\frac {1}{2} + {\sqrt {\frac {1}{4} -\bar u^2}}}{\bar u} \right )^2 +\frac {1}{\pi s} \frac {1}{{\sqrt {\frac {1}{4}-\bar u^2}}} -\frac {2-\frac {L}{2}}{s} \delta (\bar u) \, , \nonumber \\
F_0(\bar u s) &=& -2s \left [ \arcsin 2\bar u  + \bar u \ln \left ( \frac {\frac {1}{2} + {\sqrt {\frac {1}{4} -\bar u^2}}}{\bar u} \right )^2 \right ]-2\arcsin 2\bar u +\nonumber \\
&+& \pi  \left (2-\frac {L}{2}\right ) {\mbox {sgn}}\bar u \, . \label {Fexpans}
\end{eqnarray}

\subsection{The leading order density equation from the NLIE}

Once we have determined the functions $F_0(u)$ and $G_0(u,v)$, as
well as the value of the extreme $b_0$, one can write, according to
the general formula (\ref {nlin}) the nonlinear integral equation on
a finite interval satisfied by the counting function of the
$XXX_{-1/2}$ spin chain. It is not difficult to relate such NLIE
with the linear equation \cite {ES} satisfied by the density of
roots in the limit $s\rightarrow +\infty$. In this respect, it is
convenient to use the first equation in the chain (\ref {2eq1})
\begin{eqnarray}
Z_0(u)&=&-2L \arctan 2u + \int _{-b_0}^{b_0} \frac {dv}{2\pi} 2 \arctan (u-v) \frac {d}{dv}[Z_0(v) -2 L_0(v)] +\nonumber \\
&+& \sum _{h=1}^{L-2} 2 \arctan (u-u_h^{(i)}) \, , \label {Z0eq}
\end{eqnarray}
where $L_0(v)={\mbox {Im}}\ln [1+e^{iZ_0(v-i0)}]$. As before, we
rescale the variable $u=\bar u s$ and the extreme of the interval
$b_0=\bar b _0s$ and then we let $s \rightarrow +\infty$. Using the
asymptotic expansion
\begin{equation}
\arctan \bar u s = \frac {\pi}{2} {\mbox {sgn}} \bar u - \frac
{1}{\bar u s} + O(s^{-3}) \label {arctgexp}
\end{equation}
{\it inside the integral} and the expansion (\ref {bexp}) for $\bar b_0$, we obtain, {\it in this approximation}, the equation
\begin{eqnarray}
Z_0(\bar u s)&=&-\pi L {\mbox {sgn}} \bar u +\int _{-\frac
{1}{2}}^{\frac {1}{2}} \frac {d\bar v}{2\pi} \left [ \pi {\mbox
{sgn}} (\bar u -\bar v)-
\frac {2}{s(\bar u -\bar v)} \right ] \frac {d}{d\bar v}[Z_0(\bar v s)-2L_0(\bar v s)] +\nonumber \\
&+& \sum _{h=1}^{L-2} \pi {\mbox {sgn}} (\bar u-\bar u_h^{(i)}) \, .
\end{eqnarray}
Performing the integral involving the sgn function and using the
fact that, when $s\rightarrow +\infty$, $\bar u_h^{(i)} \rightarrow
0$, we are left with
\begin{equation}
2L_0(\bar u s)=-2\pi  {\mbox {sgn}} \bar u +\int _{-\frac
{1}{2}}^{\frac {1}{2}} \frac {d\bar v}{2\pi} \left [ - \frac
{2}{s(\bar u -\bar v)}\right ] \frac {d}{d\bar v}[Z_0(\bar v
s)-2L_0(\bar v s)]  \, .
\end{equation}
As the term involving $L_0(\bar vs)$ approaches zero (cf also the Appendix), we can neglect
it and obtain
\begin{equation}
0=-2 \pi  {\mbox {sgn}} \bar u - \int _{-\frac {1}{2}}^{\frac
{1}{2}} \frac {d\bar v}{2\pi} \frac {2}{s(\bar u -\bar v)} \frac
{d}{d\bar v}Z_0(\bar v s) \, .
\end{equation}
Defining the density
\begin{equation}
\bar \rho _0(\bar u) = - \frac {1}{2\pi s} \frac {d}{d\bar u}
Z_0(\bar u s) \, , \end{equation} which, because of the condition
$Z_0(b_0)-Z_0(-b_0)=-2\pi (s+L-2)$, is normalised as\begin{equation}
\int _{-\frac {1}{2}}^{\frac {1}{2}} d \bar u \bar \rho _0(\bar u)
=1 +o(s^{0}) \, ,
\end{equation}
our equation is written as,
\begin{equation}
0=-2 \pi  {\mbox {sgn}} \bar u +2 \int _{-\frac {1}{2}}^{\frac
{1}{2}} d\bar v \frac {1}{\bar u -\bar v} \bar \rho _0(\bar v) \, ,
\end{equation}
which coincides with (52) of \cite {ES}. Therefore, the NLIE on a
finite interval for the $XXX_{-1/2}$ spin chain links itself in a simple
way to the linear equation for the density of roots in the limit
$s\rightarrow +\infty$.

\subsection{Determination of $G_0$ when $s\rightarrow \infty$}

On the other hand, $G_0(u,v)$ satisfies the integral equation
\begin{equation}
G_0(u,v)=-\frac {1}{\pi} \frac {1}{1+(u-v)^2} + \int _{-b_0}^{b_0}
\frac {dw}{\pi} \frac {1}{1+(u-w)^2} G_0(w,v) \, ,
\end{equation}
which, in terms of rescaled variables reads as
\begin{equation}
G_0(\bar u s,\bar v s)=-\frac {1}{\pi} \frac {1}{1+s^2(\bar u -\bar
v)^2} + \int _{-\bar b_0}^{\bar b_0} \frac {d\bar w}{\pi} \frac
{s}{1+s^2 (\bar u-\bar w)^2} G_0(s\bar w,s\bar v) \, ,
\end{equation}
Using the expansions (\ref {functexp}), we obtain the equation
\begin{equation}
0=-\frac {1}{s} \delta (\bar u -\bar v) +\frac {1}{\pi s} \int
_{-\bar b_0}^{\bar b_0} d\bar w \left ( \frac {d}{d\bar w} P \frac
{1}{\bar u -\bar w} \right ) G_0(s\bar w, s\bar v) +o(s^{-1}) \, .
\end{equation}
We now define the order $s^0$ antisymmetric combination
\begin{equation}
X_0(\bar u, \bar v)=G_0(s\bar u , s \bar v)-G_0(s\bar u , - s \bar
v)+o(s^0) \, ,
\end{equation}
which satisfies the equation
\begin{eqnarray}
&& \delta (\bar u +\bar v)-\delta (\bar u -\bar v) + \frac {1}{\pi}
\frac {1}{\bar u -\frac {1}{2}} X_0\left (\frac {1}{2}, \bar v
\right ) -
\frac {1}{\pi} \frac {1}{\bar u +\frac {1}{2}} X_0\left (-\frac {1}{2}, \bar v \right )- \nonumber \\
&-& \frac {1}{\pi} \int _{-\frac {1}{2}}^{\frac {1}{2}} d\bar w
\left (  P \frac {1}{\bar u -\bar w} \right )\frac {d}{d\bar w}
X_0(\bar w, \bar v) =0 \, .
\end{eqnarray}
We have set $\bar b_0=1/2$, since $X_0$ takes into account only the
leading $O(s^0)$ contribution. The solution to this equation is
given by
\begin{eqnarray}
\frac {d}{d\bar u} X_0(\bar u, \bar v)&=&\left ( -\frac {1}{\pi}
\frac {1}{\bar v -\bar u}- \frac {1}{\pi} \frac {1}{\bar v +\bar
u}\right )
{\sqrt {\frac {\frac {1}{4} -\bar u^2}{\frac {1}{4} -\bar v^2}}}+ \nonumber \\
&+& 2 \left [\delta \left (\bar u -\frac {1}{2}\right ) +\delta
\left (\bar u +\frac {1}{2}\right ) \right ] X_0\left (\frac {1}{2},
\bar v \right ) \, \label{X}.
\end{eqnarray}
Integrating this function we may write down
\begin{eqnarray}
X_0(\bar u,\bar v)&=&-\frac {1}{\pi} \ln \left | \frac {\bar v+\bar
u}{\bar v -\bar u}\right| \left | \frac {{\sqrt {\frac {1}{4} -\bar
v^2}} {\sqrt {\frac {1}{4} -\bar u^2}} -\bar v\bar u +\frac {1}{4}
}{{\sqrt {\frac {1}{4} -\bar v^2}}
{\sqrt {\frac {1}{4} -\bar u^2}} +\bar v\bar u +\frac {1}{4} } \right | -\frac {2\bar v}{\pi {\sqrt {\frac {1}{4} -\bar v^2}}}\arcsin 2\bar u + \nonumber \\
&+& \left [{\mbox {sgn}}\left (\bar u -\frac {1}{2}\right )+{\mbox
{sgn}}\left (\bar u +\frac {1}{2}\right )\right ] X_0\left (\frac
{1}{2}, \bar v \right ) \, , \label {Xexpr}
\end{eqnarray}
where we took into account the fact that $X_0(u,v)=-X_0(-u,v)$ in
order to fix the undetermined function of $v$.

We also define the order $s^0$ symmetric combination
\begin{equation}
Y_0(\bar u, \bar v)=G_0(s\bar u , s \bar v)+G_0(s\bar u , - s \bar
v) +o(s^0) \, ,
\end{equation}
which satisfies the equation
\begin{equation}
-i \theta (\bar u -\bar v) + i \theta (-\bar u -\bar v)= \frac
{1}{i\pi} \int _{-\frac {1}{2}}^{\frac {1}{2}} d\bar w \left (  P
\frac {1}{\bar w -\bar u} \right ) Y_0(\bar w, \bar v) \, .
\end{equation}
This equation can be solved as
\begin{eqnarray}
Y_0(\bar u, \bar v)&=&\frac {1}{i\pi} \int _{-\frac {1}{2}}^{\frac
{1}{2}} d\bar w \left (  P \frac {1}{\bar w -\bar u} \right )
\left [ -i \theta (\bar w -\bar v)+i \theta (-\bar w -\bar v) \right ] {\sqrt {\frac {\frac {1}{4} -\bar u^2 }{\frac {1}{4} -\bar w^2 }}}= \label {Ysol} \\
&=& -\frac {1}{\pi} \int _{\bar v}^{\frac {1}{2}} d\bar w \left (  P
\frac {1}{\bar w -\bar u} \right ) {\sqrt {\frac {\frac {1}{4} -\bar
u^2 }{\frac {1}{4} -\bar w^2 }}} + \frac {1}{\pi} \int _{-\frac
{1}{2}}^{-\bar v} d\bar w \left (  P \frac {1}{\bar w -\bar u}
\right ) {\sqrt {\frac {\frac {1}{4} -\bar u^2 }{\frac {1}{4} -\bar
w^2 }}} \, . \nonumber
\end{eqnarray}
We remark that this solution satisfies the parity properties
$Y_0(\bar u, \bar v)=Y_0(\bar u, -\bar v)=Y(-\bar u, \bar v)$, as it
follows also from properties (\ref {FGparity}). As boundary
condition, we find
\begin{equation}
Y_0\left (\pm \frac {1}{2}, \bar v\right )=Y_0\left (\bar u, \pm
\frac {1}{2} \right )=0 \, .
\end{equation}

\subsection{Evaluation of the charges when $s\rightarrow \infty$}

Using (\ref {Oexp}) we want to compute the eigenvalues of the energy
and of all the charges when $s\rightarrow \infty$. We are interested
in the leading terms in the large $s$ expansion, i.e. in the terms
proportional to $\ln s$ and to $s^0$. We have to specialise formula
(\ref {Oexp}) to the case in which
\begin{eqnarray}
&&A=-b_0  \, , \quad B=b_0  \, , \quad Z_0(-b_0 )=-Z_0(b_0)=\pi (s+L-2)  \, , \\
&& O(v)=q_r(v)=\frac {i}{r-1}\left [ \frac {1}{\left (\frac
{i}{2}+v\right)^{r-1}}- \frac {1}{\left (-\frac
{i}{2}+v\right)^{r-1}} \right ] \, .
\end{eqnarray}
For parity reasons, the eigenvalues of the charges $Q_r$, with $r$
odd, are zero. Therefore, we restrict to even $r$. One easily sees
also that the first two lines of (\ref {Oexp}) give contributions at
most $O(s^{-1})$. More importantly, we have strong evidence from
numerical simulations that the nonlinear terms in the fourth line of
(\ref {Oexp}) go to zero as $s\rightarrow \infty$. This peculiar
behaviour of the nonlinear terms is due to the fact (see for example
the discussion in Appendix A of \cite {FRS}) that in our approach we
are integrating only on the interval in which the Bethe roots and
the internal holes are present and not - as in the approach of \cite
{FMQR} used by \cite {FRS} - on the whole real line. It follows
that, differently from \cite {FRS}, at least up to the order $o(s^0)$ at which
non-linear terms in the fourth line of (\ref {Oexp}) start
contributing, the eigenvalues of the charges are given by only the
linear term in the third line (cf also Appendix)
\begin{eqnarray}
Q_r&=& \int _{-b _0}^{b_0} \frac {dv}{2\pi} \frac {d}{dv} q_r(v) F_0(v)-\sum _{h=1}^{H_i}q_r(u_h^{(i)})+o(s^0)=\nonumber \\
&=&- \int _{-b _0}^{b_0} \frac {dv}{2\pi} q_r(v) \frac {d}{dv}
F_0(v)-\frac {(-1)^{1-\frac {r}{2}}}{r-1}(L-2)2^r+o(s^0) \, ,
\end{eqnarray}
which involves $F_0$, i.e. the solution of the linear integral
equation (\ref {F1loop2}).

One could now insert for $F_0$ the solution at large $s$ given by
formula (\ref {dersol}). However, we will now show that this
procedure is {\bf accurate only for the energy ($r=2$) and only for the
(coefficient of the) leading $\ln s$ term}. Indeed, let us use (\ref
{dersol}). In fact, we obtain {\it in this approximation}
\begin{eqnarray}
Q_r&=& \int _{-b_0}^{b_0} \frac {dv}{2\pi} \frac {2i}{r-1}\left [
\frac {1}{\left (\frac {i}{2}+v\right)^{r-1}}-\frac {1}{\left
(-\frac {i}{2}+v\right)^{r-1}} \right ] \Bigl [ \ln \left ( \frac
{b_0+
\sqrt {b^2_0-v^2} }{v}\right )^2-\nonumber \\
&-& \pi \left (2-\frac {L}{2}\right) \delta (v) \Bigr] -\frac {(-1)^{1-\frac {r}{2}}}{r-1}(L-2)2^r = \nonumber\\
&=& \int _{-\frac {1}{2}}^{\frac {1}{2}} dx \frac {4b_0i}{(r-1)}
\frac {1}{\left (2b_0 x+\frac {i}{2}\right ) ^{r-1}} \bar \rho _0(x)
- \frac {2^r(-1)^{1-\frac {r}{2}}}{(r-1)}\frac {L}{2} \, ,
\end{eqnarray}
identifying the Korchemsky density of Bethe roots ($s\rightarrow \infty$) \cite{Kor} as
\begin{equation}
\bar \rho _0(x)=\frac {1}{\pi} \ln \left (\frac {\frac {1}{2}+\sqrt
{\frac {1}{4}-x^2}}{x}\right )^2 \, .
\end{equation}
Introducing the resolvent
\begin{equation}
G(x)=\int _{-\frac {1}{2}}^{\frac {1}{2}}dy \frac {\bar \rho
_0(y)}{y-x} = i \ln \frac {\sqrt {1-4x^2}+1}{\sqrt {1-4x^2}-1} \, ,
\end{equation}
we see that
\begin{equation}
Q_r=\frac {2}{i} \frac {(2b_0)^{2-r}}{(r-1)!} \frac {d^{r-2}}{dx
^{r-2}} G(x) |_{x=i/4b_0} - \frac {2^r(-1)^{1-\frac
{r}{2}}}{(r-1)}\frac {L}{2} +o(s^0) \, .
\end{equation}
Explicit computation would give, for $r=2$,
\begin{equation}
Q_2=E= 4\ln s + 4\ln 2 -2L + o(s^0) \, ,
\end{equation}
and, for $r\geq 4$,
\begin{equation}
Q_r=\frac {2^r (-1)^{1-\frac {r}{2}}}{(r-1)(r-2)} - \frac
{2^r(-1)^{1-\frac {r}{2}}}{(r-1)}\frac {L}{2}+o(s^0)= \frac {2^r
(-1)^{1-\frac {r}{2}}}{(r-1)(r-2)} \left [ 1-\frac {L}{2}(r-2)
\right ]+o(s^0)\, .
\end{equation}
One can check for $L=2,3$ that the order $s^0$ terms in the energy
and in the higher charges are {\bf not the correct ones} (cf. for instance
\cite{Kor,ES,Bec,KLRSV}). Therefore, the large spin solution for
$F_0$, (\ref {dersol}), is a {\it good} approximation only if we are
interested in the leading $O(\ln s)$ term of the energy, although
$F_0$ would be {\it exact} up to the $O(s^0)$ order, i.e. neglecting
(the going to zero) $o(s^0)$ terms. Yet, this failure reflects the
subtlety of considering the $s\rightarrow +\infty$ limit of the
integral equation (\ref{F1loop}).

Nevertheless, a sufficiently accurate approximation for $F_0$,
efficient when neglecting in the charges the $o(s^0)$ terms
\footnote{More precisely, the first order we are neglecting is
$O(1/\ln s)$ and comes from the approximation of all the internal
hole positions $u^{(i)}_h=0$, which is strictly valid only if
$s=+\infty$.}, comes out by solving the one loop density equation,
i.e. the derivative of (both members of) equation (\ref{F1loop}),
upon approximating $u^{(i)}_h=0$, by means of Fourier transform
technique
\begin{equation}
ik\hat F_0(k) = -4\pi \frac {\frac {L}{2}-e^{-\frac {|k|}{2}}\cos (k
s/\sqrt{2})} {2\sinh \frac {|k|}{2}}+2\pi (L-2) \frac {e^{-\frac
{|k|}{2}}}{2\sinh \frac {|k|}{2}}-(4\pi \ln 2) \delta (k) + o(s^0) \,
. \label {Fkappa}
\end{equation}
Using (\ref {Fkappa}) and the Fourier transform of the charges
functions $\hat q_r(k)$, the eigenvalues of $Q_r$ at order $\ln s$
and $s^0$ are given by
\begin{equation}
Q_r=-\int _{-\infty}^{+\infty} \frac {dk}{4\pi^2} \hat q_r(k) ik
\hat F_0(k) -\frac {(-1)^{1-\frac {r}{2}}}{r-1}(L-2)2^r+o(s^0)\, .
\label{Ealpha}
\end{equation}
From (\ref {Ealpha}) we obtain the correct results for the eigenvalues
of the charges
\begin{eqnarray}
Q_2&=&E=4\ln s + 4 \gamma _E -4(L-2)\ln 2 +o(s^0)\, , \nonumber \\
Q_r&=&\frac {2 (-1)^{1-\frac {r}{2}}\zeta
(r-1)}{r-1}[(2-2^{r-1})L-2(1-2^{r-1})] + o(s^0) \, , \quad r \geq 4
\, ,
\end{eqnarray}
where $\gamma _E$ is the Euler-Mascheroni constant and $\zeta (x)$
is the Riemann zeta function.

\section {All loops}
\setcounter{equation}{0}

Let us now consider the Bethe ansatz like equations
\begin{equation}
\left ( \frac {u_k+\frac {i}{2}}{u_k-\frac {i}{2}} \right )^L \left
( \frac {1+\frac {g^2}{2{x_k^-}^2}}{1+\frac {g^2}{2{x_k^+}^2}}
\right )^L=\mathop{\prod^s_{j=1}}_{j\neq k}  \frac
{u_k-u_j-i}{u_k-u_j+i}  \left ( \frac {1-\frac
{g^2}{2x_k^+x_j^-}}{1-\frac {g^2}{2x_k^-x_j^+}} \right )^2
e^{2i\theta (u_k,u_j)}\, , \label {manyeq}
\end{equation}
where we used the notations
\begin{equation}
x^{\pm}_k=x^{\pm}(u_k)=x(u_k\pm i/2) \, , \quad x(u)=\frac
{u}{2}\left [ 1+{ \sqrt {1-\frac {2g^2}{u^2}}} \right ] \, , \quad
\lambda =8\pi ^2 g^2 \, ,
\end{equation}
$\lambda $ being the 't Hooft coupling. It is believed that
configurations of Bethe roots, i.e. solutions of (\ref {manyeq}),
and the corresponding eigenvalues of the energy are related
respectively to composite operators and their anomalous dimensions
in the $sl(2)$ sector of ${\cal N}=4$ SYM. This correspondence,
however, breaks at the so-called wrapping order, i.e. at order
$g^{2L-2}$, and higher. Therefore, results in this section are
relevant for ${\cal N}=4$ SYM only until the order $g^{2L-4}$.

In a fashion similar to the one loop case, Bethe roots concentrate
in an interval $[-b,b]$ of the real axis. Inside this interval,
$L-2$ holes are present, while outside it two external holes lie. We
use for them the same notations as in the one loop case.

The counting function is
\begin{eqnarray}
Z(u)&=&-2L \arctan 2u -i L \ln \left ( \frac {1+\frac {g^2}{2{x^-(u)}^2}}{1+\frac {g^2}{2{x^+(u)}^2}} \right )-2\sum _{j=1}^s \arctan (u-u_j) + \nonumber \\
&+& 2i \sum _{j=1}^s \ln \left ( \frac {1-\frac
{g^2}{2x^+(u)x_j^-}}{1-\frac {g^2}{2x^-(u)x_j^+}} \right ) -2\sum
_{j=1}^s \theta (u,u_j) \,   , \label {nliemany}
\end{eqnarray}
where the so-called dressing factor is given by
\begin{equation}
\theta (u_k,u_j)=\sum _{r=2}^{\infty}\sum _{\nu =0}^{\infty} \beta
_{r,r+1+2\nu}(g)
[q_r(u_k)q_{r+1+2\nu}(u_j)-q_r(u_j)q_{r+1+2\nu}(u_k)] \, ,
\end{equation}
the functions $\beta _{r,r+1+2\nu}(g)$ being
\begin{eqnarray}
\beta _{r,r+1+2\nu}(g)&=&2 \sum _{\mu =\nu}^{\infty} \frac {g^{2r+2\nu+2\mu}}{2^{r+\mu+\nu}} (-1)^{r+\mu+1}\frac {(r-1)(r+2\nu)}{2\mu +1} \cdot \nonumber \\
&\cdot & \left ( \begin{array}{cc} 2\mu +1 \\ \mu -r-\nu+1
\end{array} \right )\left ( \begin{array}{cc} 2\mu +1 \\ \mu -\nu
\end{array} \right )              \zeta (2\mu +1)
\end{eqnarray}
and $q_r(u)$ being the density of the $r$-th charge
\begin{equation}
q_r(u)=\frac {i}{r-1} \left [ \left (\frac {1}{x^+(u)}\right
)^{r-1}-\left (\frac {1}{x^-(u)}\right )^{r-1} \right ] \, .
\end{equation}

\subsection {Equations for $F$, $F_0$ and $F^H$.}

The counting function (\ref {nliemany}) can be treated according to
the general formalism given in Section 2, provided we choose $A=-b$,
$B=b$ and
\begin{eqnarray}
\Phi (u) &=&-2L \arctan 2u -iL \ln \left ( \frac {1+\frac {g^2}{2{x^-(u)}^2}}{1+\frac {g^2}{2{x^+(u)}^2}} \right )\, ,  \nonumber \\
\phi (u,v)&=& 2\arctan (u-v) -2i \left [ \ln \left ( \frac {1-\frac {g^2}{2x^+(u)x^-(v)} }{1-\frac {g^2}{2x^-(u)x^+(v)}} \right )+i\theta (u,v)\right] \, , \nonumber \\
f(u)&=&-2L \arctan 2u -iL \ln \left ( \frac {1+\frac {g^2}{2{x^-(u)}^2}}{1+\frac {g^2}{2{x^+(u)}^2}} \right )  + \nonumber \\
&+& 2\sum _{h=1}^{L-2} \left [ \arctan (u-u_h^{(i)}(g))-i \ln \left (\frac {1-\frac {g^2}{2x^+(u)x^-(u_h^{(i)}(g))}}{1-\frac {g^2}{2x^-(u)x^+(u_h^{(i)}(g))}} \right )+\theta (u,u_h^{(i)}(g))\right] \, , \nonumber \\
&+& \frac {1}{\pi} Z(b) \Bigl \{ \left [ \arctan (u-b)-i \ln \left ( \frac {1-\frac {g^2}{2x^+(u)x^-(b)} }{1-\frac {g^2}{2x^-(u)x^+(b)}} \right )+\theta (u,b) \right ]  +  \nonumber \\
&+& \left [ \arctan (u+b)-i \ln \left ( \frac {1-\frac
{g^2}{2x^+(u)x^-(-b)} }{1-\frac {g^2}{2x^-(u)x^+(-b)}} \right )
+\theta (u,-b)\right ] \Bigr \} \, , \label {ESfun}
\end{eqnarray}
with the explicit $g$-dependence of the internal all loop holes
$u_h^{(i)}(g)$. We have supposed that ${\mbox {Im}}\ln [1+e^{iZ(\pm
b)}]=0$, as in the one loop case. It follows that the function
$F(u)$ entering the NLIE satisfies the linear equation (\ref {Feq})
with the function $f(u)$ and $\varphi (u,v)=\frac {1}{2\pi} \frac
{d}{dv} \phi (u,v)$ obtained from (\ref {ESfun}). We can split
$F(u)$ into its one loop contribution $F_0$ and its higher loop
contribution $F^{H}(u)$:
\begin{equation}
F(u)=F_0(u)+F^{H}(u) \, .
\end{equation}
Of course, the one loop contribution satisfies the LIE
(\ref{F1loop})
\begin{eqnarray}
F_0(u)&=& - 2L \arctan 2u  - (s+L-2) [\arctan (u-b_0)+\arctan (u+b_0) ]  + \nonumber \\
&+& 2\sum _{h=1}^{L-2} \arctan (u-u_h^{(i)}(0))+\int _{-b_0} ^{b_0}
dv \frac {1}{\pi} \frac {1}{1+( u -v)^2} F_0(v)  \, ,
\end{eqnarray}
where $u_h^{(i)}(0)$ are indeed the internal one loop holes (as
$g=0$ value of the (internal) all loop holes). On the contrary, the
LIE obeyed by the higher loop $F^{H}(u)$ contains additionally the
$g$-depending holes in the form
\begin{eqnarray}
&&F^{H}(u)= -iL \ln \left ( \frac {1+\frac {g^2}{2{x^-(u)}^2}}{1+\frac {g^2}{2{x^+(u)}^2}} \right ) -2i \sum _{h=1}^{L-2}\Bigl [ \ln \left ( \frac {1-\frac {g^2}{2x^+(u)x^-(u_h^{(i)}(g))} }{1-\frac {g^2}{2x^-(u)x^+(u_h^{(i)}(g))}} \right )+i\theta (u,u_h^{(i)}(g))+\nonumber \\
&+& i \arctan (u-u_h^{(i)}(g))-i\arctan (u-u_h^{(i)}(0))  \Bigr ] - \nonumber \\
&-&\frac {i}{\pi}Z(b) \left [ \ln \left ( \frac {1-\frac
{g^2}{2x^+(u)x^-(b)} }{1-\frac {g^2}{2x^-(u)x^+(b)}} \right ) +
\ln \left ( \frac {1-\frac {g^2}{2x^+(u)x^-(-b)} }{1-\frac {g^2}{2x^-(u)x^+(-b)}} \right )+i\theta (u,b)+i\theta (u,-b) \right ]  + \nonumber \\
&+&\frac {1}{\pi}Z(b) [\arctan (u-b)+\arctan (u+b)]-
\frac {1}{\pi}Z(b_0) [\arctan (u-b_0)+\arctan (u+b_0)] + \nonumber \\
&+& \int _{-b}^b \frac {dv}{\pi} \frac {1}{1+(u-v)^2}F^{H}(v) + \int
_{-b}^{-b_0} \frac {dv}{\pi} \frac {1}{1+(u-v)^2}F_{0}(v) +
\int _{b_0}^{b} \frac {dv}{\pi} \frac {1}{1+(u-v)^2}F_{0}(v) + \nonumber \\
&+& \frac {i}{\pi} \int _{-b}^b dv \left [ \frac {d}{dv} \ln \left (
\frac {1-\frac {g^2}{2x^+(u)x^-(v)} }{1-\frac {g^2}{2x^-(u)x^+(v)}}
\right )+i \frac {d}{dv} \theta (u,v) \right ] [F_0(v)+F^{H}(v)] \,
. \label {Ftilde}
\end{eqnarray}

Now, we proceed to the limit $s\rightarrow +\infty$ keeping, of
course, only the non-vanishing terms. Upon treating by parts the
last integral, the term in the square brackets in the third line of
(\ref {Ftilde}) gets multiplied by the factor
$Z(b)-F(b)=o(s^0)$\footnote {This is certainly true at one loop, as
follows by comparing (\ref {Zcond}) and (\ref {Fcond}). We suppose
that it stays true also when considering all the loop corrections.}.
Therefore it can be neglected. Furthermore, since
$u_h^{(i)}(g)=o(s^0)$, we can set everywhere $u_h^{(i)}(g)=0$.
Moreover, as $s\rightarrow +\infty$, one has that $Z(b)=O(s)$,
$Z_0(b_0)=O(s)$, but also that $\arctan (u-b)+\arctan (u+b)=
O(s^{-2})$,  $\arctan (u-b_0)+\arctan (u+b_0)= O(s^{-2})$. So, the
terms in the fourth line can be neglected as well. Since $b-b_0$ is
$O(s^0)$ and $F_0(b)$ is $O(s)$, the second and the third integral
in the fifth line can also be removed. For the same reason, in the
integrals in the sixth line involving $F_0$ we can set the extremes
to their one loop value $\mp b_0$. Finally, in the integrals
involving $F^H(v)$ we may replace the extreme, $b$, -- which in
principle ia an unknown of the integral equation (\ref{Ftilde}) as
well -- with $+\infty$, since as $v\rightarrow \infty$ $F^H(v)$
becomes constant, while the rest of the integrands vanishes as
$1/v^2$ or faster. Rearranging the terms in the right hand side, we have
\begin{eqnarray}
F^{H}(u)&=& -iL \ln \left ( \frac {1+\frac
{g^2}{2{x^-(u)}^2}}{1+\frac {g^2}{2{x^+(u)}^2}} \right )
-2i(L-2)\left [ \ln \left ( \frac {1-\frac
{g^2}{2x^+(u)x^-(0)}}{1-\frac {g^2}{2x^-(u)x^+(0)}}\right ) + i
\theta (u,0) \right ]-
\nonumber \\
&-&  \frac {i}{\pi} \int _{-b_0}^{b_0} dv \left [ \ln \left ( \frac
{1-\frac {g^2}{2x^+(u)x^-(v)} }{1-\frac {g^2}{2x^-(u)x^+(v)}} \right
)+i \theta (u,v) \right ]
\frac {d}{dv} F_0(v)+ \nonumber \\
&+&  \int _{-\infty}^{+\infty} \frac {dv}{\pi} \frac {1}{1+(u-v)^2}F^{H}(v) - \label {Ftildes} \\
&-& \frac {i}{\pi} \int _{-\infty}^{+\infty} dv \left [ \ln \left (
\frac {1-\frac {g^2}{2x^+(u)x^-(v)} }{1-\frac {g^2}{2x^-(u)x^+(v)}}
\right )+i \theta (u,v) \right ] \frac {d}{dv} F^{H}(v) +o(s^0)\, .
\nonumber
\end{eqnarray}
With such rearrangement, we have collected in the first two lines of
the right hand side the forcing terms of our equation. In the last
two lines, we have the integral terms, i.e. terms involving integral
of a kernel function with the unknown $F^H$.

Several comments are now in order. The LIE (\ref{Ftildes})
constraining the density of roots $-\frac{1}{2\pi}\frac
{d}{du}F^H(u)$ may be thought of as an improvement of the BES
equation \cite{BES}, which in fact takes into account not only the
leading $O(\ln s)$ term, but also the subleading $O(s^0)$ (constant)
corrections. This equation is exact and linear and therefore drives
the interpolation between weak (small $g$) to strong (large $g$)
coupling in a non-perturbative way. In particular, it may be useful
for studying the strong coupling regime where the string effects,
and in particular the dressing phase, dominate. However, in view of
possible comparisons and checks vs. string results, one should
fucus the attention in cases where the aforementioned wrapping
effects are negligible or known.

As in the one loop case, in order to compute the energy and the
eigenvalues of the charges in the limit $s\rightarrow \infty$ and including 
constant terms, in our framework it is sufficient to
consider the third line of the general formula (\ref {Oexp}).
Explicitly,
\begin{eqnarray}
E(g,s)&=&\int _{-b_0}^{b_0}\frac {dv}{2\pi}  \frac {d}{dv} \left
[\frac {i}{x^+(v)}-\frac {i}{x^-(v)}\right ] F_0(v)-(L-2)
 \left [\frac {i}{x^+(0)}-\frac {i}{x^-(0)}\right ]+\nonumber \\
&+& \int _{-\infty}^{+\infty}\frac {dv}{2\pi}  \frac {d}{dv} \left
[\frac {i}{x^+(v)}-\frac {i}{x^-(v)}\right ] F^H(v) + o(s^{0})\, .
\label{egs}
\end{eqnarray}
This means that, in contrast to the approach of \cite {FRS}, where
the \cite {FMQR} method is used, we have to cope only with {\it
linear} equations. In respect to (\ref {egs}) and for future
convenience, we introduce the function
\begin{eqnarray}
h(g,s)&=&-\int _{-b_0}^{b_0}\frac {dv}{2\pi}  \left [\frac
{i}{x^+(v)}-\frac {i}{x^-(v)}\right ]  \frac {d}{dv} F_0(v)-(L-2)
 \left [\frac {i}{x^+(0)}-\frac {i}{x^-(0)}\right ] = \nonumber \\
&=&\int _{-b_0}^{b_0}\frac {dv}{2\pi} \frac {2i}{x^-(v)} \frac
{d}{dv} F_0(v)+2i (L-2)\frac {1}{x^-(0)} \, ,
\end{eqnarray}
in such a way that the energy reads
\begin{equation}
E(g,s)=h(g,s)-\int _{-\infty}^{+\infty}\frac {dv}{2\pi}  \left
[\frac {i}{x^+(v)}-\frac {i}{x^-(v)}\right ] \frac {d}{dv}
F^{H}(v)+o(s^0) \, .
\end{equation}
Direct calculations show that
\begin{eqnarray}
h(g,s)&=&4\ln s +4\gamma _E -4(L-2)\ln 2 +h(g) \, , \label {hgs} \\
h(g)&=&2g^2[3L\zeta (3)-7\zeta (3)] + g^4(62-30L)\zeta (5)+ O(g^6)
+o(s^0) \, . \nonumber
\end{eqnarray}
In addition, we remark that equation (\ref {Ftildes}) has the same
kernel as the BES equation. They differ only in their forcing terms,
since in the BES case, in contrast to (\ref {Ftildes}), the forcing
terms are simply
\begin{eqnarray}
&&\left \{ \frac {ig^2}{2\pi}\left [ \frac {1}{x^+(u)}+\frac {1}{x^-(u)}\right ]+\frac {2i}{\pi}\sum _{\nu =0}^{\infty}\beta _{2,3+2\nu}(g)q_{3+2\nu}(u) \right \} \int _{-b_0}^{b_0} dv \frac {1}{v-\frac {i}{2}} \frac {d}{dv} F_0(v)= \nonumber \\
&=&2g^2 \ln s \left [ \frac {1}{x^+(u)}+\frac {1}{x^-(u)}\right ]+
8\ln s \sum _{\nu =0}^{\infty}\beta _{2,3+2\nu}(g)q_{3+2\nu}(u) +
O(s^0) \, , \label {BES}
\end{eqnarray}
where we used the one loop results at order $\ln s$:
\begin{equation}
\int _{-b_0}^{b_0} \frac {dv}{2\pi} \frac {1}{v+\frac {i}{2}} \frac
{d}{dv} F_0(v)=-\int _{-b_0}^{b_0} \frac {dv}{2\pi} \frac
{1}{v-\frac {i}{2}} \frac {d}{dv} F_0(v) =2i\ln s + O(s^0) \, .
\end{equation}
Our forcing terms differ from the ones in the BES equation for two
reasons. Firstly, in addition to the BES terms, we have genuine new
terms coming from the first line of the right hand side of (\ref
{Ftildes}). If we expand them in powers of $g^2$, we see that they
are structurally different from the BES forcing terms, with the
exception of the term
\begin{eqnarray}
&&2i(L-2)\left [ \frac {g^2}{2x^+(u)x^-(0)}-\frac {g^2}{2x^-(u)x^+(0)}\right ] -2(L-2)\sum _{\nu =0}^{\infty}\beta _{2,3+2\nu}(g)q_{3+2\nu}(u)q_2(0) = \nonumber \\
&& \left \{ \frac {ig^2}{2\pi}\left [ \frac {1}{x^+(u)}+\frac
{1}{x^-(u)}\right ]+\frac {2i}{\pi}\sum _{\nu =0}^{\infty}\beta
_{2,3+2\nu}(g)q_{3+2\nu}(u) \right \}\frac {2\pi (L-2)}{x^-(0)} \, ,
\end{eqnarray}
which is proportional to the BES forcing terms. Secondly, for what
concerns the terms in the second line of the right hand side of
(\ref {Ftildes}), in their expansions in powers of $g^2$ all the
terms have to be kept, since we want to be precise in $s$ up to the
order $s^0$. However, some of these terms, namely
\begin{eqnarray}
\left \{ \frac {ig^2}{2\pi}\left [ \frac {1}{x^+(u)}+\frac
{1}{x^-(u)}\right ]+\frac {2i}{\pi}\sum _{\nu =0}^{\infty}\beta
_{2,3+2\nu}(g)q_{3+2\nu}(u) \right \} \int _{-b_0}^{b_0} dv \frac
{1}{x^-(v)} \frac {d}{dv} F_0(v) \, , \nonumber
\end{eqnarray}
as functions of $u$, are also proportional to the BES ones. We
conclude that in our equation (\ref {Ftildes}) a part of the forcing
terms,
\begin{eqnarray}
&&\left \{ \frac {ig^2}{2\pi}\left [ \frac {1}{x^+(u)}+\frac {1}{x^-(u)}\right ]+\frac {2i}{\pi}\sum _{\nu =0}^{\infty}\beta _{2,3+2\nu}(g)q_{3+2\nu}(u) \right \} \cdot \nonumber \\
&\cdot & \left [ \int _{-b_0}^{b_0} dv \frac {1}{x^-(v)} \frac
{d}{dv} F_0(v)+
\frac {2\pi (L-2)}{x^-(0)}\right ] = \nonumber \\
&=& \left \{ \frac {ig^2}{2\pi}\left [ \frac {1}{x^+(u)}+\frac
{1}{x^-(u)}\right ]+\frac {2i}{\pi}\sum _{\nu =0}^{\infty}\beta
_{2,3+2\nu}(g)q_{3+2\nu}(u) \right \} \frac {\pi}{i} h(g,s) \, ,
\label {BESpro}
\end{eqnarray}
is proportional to the forcing terms in the BES equation (the
forcing term in the BES equation is given by (\ref {BESpro}) in
which $h(g,s)$ is replaced by $4\ln s$). Therefore we are allowed to
say that the solution of our equation is the sum of a function
proportional to the solution of the BES equation and an unknown
function,
\begin{equation}
s \rightarrow +\infty \, , \quad F^H(u)=\frac {h(g,s)}{4\ln s}
F^{BES}(u)+F^{extra}(u) + o(s^0) \, .
\end{equation}
This means that in the expression for the energy $E(g,s)$ we expect
that
\begin{eqnarray}
E(g,s)&=&h(g,s)+\frac {1}{4}h(g,s)[f(g)-4]+E^{extra}(g,s) +o(s^0)=\nonumber \\
&=&\frac {1}{4}h(g,s)f(g)+E^{extra}(g,s) +o(s^0) \, , \label{Estru}
\end{eqnarray}
where $f(g)$ is the universal scaling function of ${\cal N}=4$ SYM
and $E^{extra}(g,s)$ indicates contributions coming from
$F^{extra}(u)$.

\medskip

We can find the structure (\ref {Estru}) for the energy $E(g,s)$, by
performing a {\it brute force} perturbative expansion in order to
solve equation (\ref {Ftildes}) and to compute the energy. Let us
define
\begin{equation}
F^{H}(u)=g^2 F_1^{H}(u)+g^4 F_2^{H}(u) + O(g^6) \, . \label{bf}
\end{equation}
At the order $g^2$ we have the equation
\begin{eqnarray}
F_1^{H}(u)&=&\frac {L}{2i}\left [ \frac {1}{\left (u-\frac {i}{2}\right)^2}- \frac {1}{\left (u+\frac {i}{2}\right)^2} \right]+ \int _{-b}^b \frac {dv}{\pi} \frac {1}{1+(u-v)^2}F_1^{H}(v) -\nonumber \\
&-&2(L-2) \left (\frac {1}{u+\frac {i}{2}}+\frac {1}{u-\frac
{i}{2}}\right)
+\frac {1}{2i\pi}\frac {1}{u-\frac {i}{2}}\int _{-b}^b dv \frac {1}{v+\frac {i}{2}} \frac {d}{dv} F_0(v) -\nonumber \\
&-&\frac {1}{2i\pi}\frac {1}{u+\frac {i}{2}}\int _{-b}^b dv \frac
{1}{v-\frac {i}{2}} \frac {d}{dv} F_0(v) +o(s^0) \, .
\end{eqnarray}
We can now use the one loop results and then pass to the Fourier
transform
\begin{eqnarray}
\hat F_1^{H}(k)&=&\frac {\pi L  k}{i}e^{-\frac {|k|}{2}}+e^{-|k|}\hat F_1^{H}(k)-\nonumber \\
&-& 4\pi i [\ln s + \gamma _E-(L-2)\ln 2] {\mbox {sgn}}(k)e^{-\frac
{|k|}{2}} \label{FH1}
\end{eqnarray}
Solving this equation and going back to the coordinate space we reach
\begin{equation}
F_1^{H}(u)=2\pi   [\ln s + \gamma _E-(L-2)\ln 2]\tanh u \pi +\frac
{L}{2i} \left [ \psi ^{\prime}\left (\frac {1}{2}-iu\right)-\psi
^{\prime}\left (\frac {1}{2}+iu\right) \right ] \, . \nonumber
\end{equation}
In a similar fashion, one computes $F_2^{H}(u)$. We omit the details
and give only the final result:
\begin{eqnarray}
F_2^{H}(u)&=&\frac {L}{16}\frac {d^3}{du^3}\left [ \psi \left (\frac {1}{2}-iu\right )+ \psi \left (\frac {1}{2}+iu\right ) \right ] - \nonumber \\
&-& \frac {\pi ^2}{12}(L-3)
\frac {d}{du} \left [ \psi \left (\frac {1}{2}-iu\right )+ \psi \left (\frac {1}{2}+iu\right ) \right ] + \nonumber \\
&+& \frac {\pi}{2}[ \ln s + \gamma _E -(L-2)\ln 2]\frac
{d^2}{du^2}\tanh \pi u
- \nonumber \\
&-& \pi \left \{ \frac {\pi ^2}{3}[ \ln s + \gamma _E -(L-2)\ln 2]+7
\zeta (3)-2L \zeta (3) \right \} \tanh \pi u \, . \label{FH2}
\end{eqnarray}
Therefore, we are allowed to write that
\begin{eqnarray}
F^H(u)&=&[ \ln s + \gamma _E -(L-2)\ln 2]\left ( 2\pi g^2 \tanh \pi u +\frac {\pi}{2}  g^4 \frac {d^2}{du^2}\tanh \pi u - \frac {\pi ^3}{3}g^4  \tanh \pi u \right ) - \nonumber \\
&-& \pi (7 \zeta (3)-2L \zeta (3) ) g^4 \tanh \pi u+ \nonumber \\
&+& \left [ \frac {L}{2}g^2-\frac {\pi ^2}{12}(L-3)g^4 \right ] \frac {d}{du} \left [ \psi \left (\frac {1}{2}-iu\right )+ \psi \left (\frac {1}{2}+iu\right ) \right ] + \nonumber \\
&+&g^4 \frac {L}{16}\frac {d^3}{du^3}\left [ \psi \left (\frac
{1}{2}-iu\right )+ \psi \left (\frac {1}{2}+iu\right ) \right
]+O(g^6) \, .
\end{eqnarray}
After obtaining these results we can evaluate the energy $E(g,s)$ up
to the order $g^4$ and at the orders $\ln s$ and $s^0$. From the
general formula (\ref {Oexp}) we may write that
\begin{equation}
E(g,s)=\int _{-b}^{b}\frac {dv}{2\pi}  \frac {d}{dv} \left [\frac
{i}{x^+(v)}-\frac {i}{x^-(v)}\right ] [F_0(v)+F^{H}(v)]-(L-2)
 \left [\frac {i}{x^+(0)}-\frac {i}{x^-(0)}\right ]+o(s^{0}) \, .
\end{equation}
Defining the coefficients of the expansion in powers of $g^2$ as
\begin{equation}
E(g,s)=E_0(s)+E_1(s)g^2 +E_2(s)g^4+O(g^6) \, , \label{egsbf}
\end{equation}
we have that
\begin{eqnarray}
E_0(s)&=&\int _{-b}^{b}\frac {dv}{2\pi}  \frac {d}{dv} \left [\frac {i}{v+\frac {i}{2}}-\frac {i}{v-\frac {i}{2}}\right ] F_0(v)-4(L-2)+O(s^{-1})=\nonumber \\
&=& 4\ln s + 4\gamma _E -4(L-2) \ln 2 +o(s^{0}) \, .
\end{eqnarray}
On the other hand the order $g^2$ of $E(g,s)$ is given by
\begin{eqnarray}
E_1(s)&=&\int _{-b}^{b}\frac {dv}{2\pi}  \frac {d}{dv} \left [\frac {i}{v+\frac {i}{2}}-\frac {i}{v-\frac {i}{2}}\right ]F_1^{H}(v)+ \label {E1} \\
&+&\int _{-b}^{b}\frac {dv}{2\pi}  \frac {d}{dv} \left [\frac {\frac
{i}{2}}{\left (v+\frac {i}{2}\right )^3}-\frac {\frac {i}{2}}{\left
(v-\frac {i}{2}\right )^3}\right ] F_0(v)+8(L-2)+o(s^{0}) \, .
\nonumber
\end{eqnarray}
The first line of (\ref {E1}) gives
\begin{eqnarray}
&&\int _{-b}^{b}\frac {dv}{2\pi}  \frac {d}{dv} \left [\frac {i}{v+\frac {i}{2}}-\frac {i}{v-\frac {i}{2}}\right ]F_1^{H}(v)=-\int _{-b}^{b}\frac {dv}{2\pi}  \left [\frac {i}{v+\frac {i}{2}}-\frac {i}{v-\frac {i}{2}}\right ]\frac {d}{dv} F_1^{H}(v)+o(s^{0})=\nonumber \\
&=& -\pi  [\ln s +\gamma _E-(L-2)\ln 2] \int _{-b}^{b} dv \frac {1}{\frac {1}{4}+v^2} \frac {1}{\cosh ^2 \pi v} - \nonumber \\
&-& \int _{-b}^{b}\frac {dv}{2\pi}\left [ \frac {2i}{\left (v+\frac {i}{2}\right )^3}-\frac {2i}{\left (v-\frac {i}{2}\right )^3} \right ] \frac {L}{2} \left [ \psi \left (\frac {1}{2}-iv\right )+\psi \left (\frac {1}{2}+iv\right ) \right ]+o(s^{0})= \nonumber \\
&=&  -\frac {2}{3}\pi ^2  [\ln s +\gamma _E-(L-2)\ln 2]-2L \zeta (3)
+o(s^{0})\, .
\end{eqnarray}
On the other hand, the second line of (\ref {E1}) gives
\begin{eqnarray}
&&\int _{-b}^{b}\frac {dv}{2\pi}  \frac {d}{dv} \left [\frac {\frac {i}{2}}{\left (v+\frac {i}{2}\right )^3}-\frac {\frac {i}{2}}{\left (v-\frac {i}{2}\right )^3}\right ] F_0(v)+8(L-2) = \nonumber \\
&=&\frac {L}{2} \psi ^{''}(1)-\frac {L-2}{2}\psi ^{''}\left (\frac
{3}{2} \right )+8(L-2)+o(s^{0})=6L\zeta (3)-14 \zeta (3)+o(s^{0})\,
.\nonumber
\end{eqnarray}
Adding up the two contributions we obtain
\begin{equation}
E_1(s)=-\frac {2}{3}\pi ^2  [\ln s +\gamma _E-(L-2)\ln 2]+ 4L\zeta
(3)-14 \zeta (3)+o(s^{0}) \, . \label {Edue}
\end{equation}
Analogously the third order in the energy, $E_2(s)$, given by
\begin{eqnarray}
E_2(s)&=&\int _{-b}^{b}\frac {dv}{2\pi}  \frac {d}{dv} \left [\frac {1}{\frac {1}{2}-iv}+\frac {1}{\frac {1}{2}+iv}\right ]F_2^{H}(v)- \label {E2} \\
&-&\int _{-b}^{b}\frac {dv}{2\pi}  \frac {d}{dv} \left [\frac {\frac {1}{2}}{\left (\frac {1}{2}-iv\right )^3}+\frac {\frac {1}{2}}{\left (\frac {1}{2}+iv\right )^3}\right ] F_1^H(v)+\nonumber \\
&+&\int _{-b}^{b}\frac {dv}{2\pi}  \frac {d}{dv} \left [\frac {\frac
{1}{2}}{\left (\frac {1}{2}-iv\right )^5}+\frac {\frac {1}{2}}{\left
(\frac {1}{2}+iv\right )^5}\right ] F_0(v)-32(L-2)+o(s^{0}) \, ,
\nonumber
\end{eqnarray}
after similar and lengthy calculations, equals
\begin{eqnarray}
E_2(s)&=&\frac {11}{45} \pi ^4 [\ln s +\gamma _E -(L-2)\ln 2 ] + \nonumber \\
&+& \frac {\pi ^2}{3}(4-L)\zeta (3)+(62-21L) \zeta (5)  + o(s^0) \,
. \label {Etre}
\end{eqnarray}
Collecting all these terms we have
\begin{eqnarray}
E(g,s)&=&[\ln s +\gamma _E -(L-2)\ln 2 ]\left ( 4 -\frac {2}{3}\pi ^2 g^2 +\frac {11}{45}\pi ^4 g^4 \right ) + g^2 \left [ 4L \zeta (3)-14 \zeta (3) \right ] + \nonumber \\
&+&  g^4 \frac {\pi ^2}{3}(4-L)\zeta (3)+g^4 (62-21L) \zeta (5)
+O(g^6)+o(s^0) \, . \label {Egs3}
\end{eqnarray}
We recognize in (\ref {Egs3}) the expansion of the universal scaling
function $f(g)$,
\begin{equation}
f(g)=  4 -\frac {2}{3}\pi ^2 g^2 +\frac {11}{45}\pi ^4 g^4 + O(g^6)
\, ,
\end{equation}
and remark that, consistently with (\ref {Estru}), relation (\ref
{Egs3}) can be written also
\begin{equation}
E(g,s)=\frac {1}{4}h(g,s)f(g)+E^{extra}(g,s)+O(g^6)+o(s^0) \, ,
\end{equation}
where $h(g,s)$ is given by (\ref {hgs}) and
\begin{equation}
E^{extra}(g,s)=-2g^2 L \zeta (3)+\frac {2}{3}\pi ^2 g^4 L \zeta
(3)-\pi ^2 g^4 \zeta (3)+9Lg^4 \zeta (5)+O(g^6) \, .
\end{equation}
We remark also that expansion (\ref {Egs3}) agrees with result
(3.16) of \cite {FRS}.

\medskip

Finally, using similar techniques one computes, up to the order
$g^6$, the eigenvalues of all the conserved charges,
\begin{equation}
Q_r(g,s)=\frac {i}{r-1}\sum _{k=1}^s \left [ \left (\frac
{1}{x^+(u_k)}\right )^{r-1}-\left (\frac {1}{x^-(u_k)}\right )^{r-1}
\right ] \, ,
\end{equation}
with $r$ even, $r\geq 4$. Defining
\begin{equation}
Q_r(g,s)=Q_{r,0}(s)+Q_{r,1}(s)g^2 +Q_{r,2}(s)g^4+O(g^6) \, ,
\end{equation}
we obtain the following results
\begin{eqnarray}
Q_{r,0}(s)&=&\frac {2 (-1)^{\frac {r}{2}-1}\zeta (r-1)}{r-1}[(2-2^{r-1})L-2(1-2^{r-1})] + o(s^0) \, , \nonumber \\
\nonumber \\
Q_{r,1}(s)&=&4(-1)^{\frac {r}{2}}\zeta (r)  [\ln s +\gamma _E -(L-2)\ln 2 ]+ \nonumber \\
&+& L (-1)^{\frac {3r}{2}} (r+2-2^{r+1})\zeta (r+1)+2 (-1)^{\frac {3r}{2}}(2^{r+1}-1)\zeta (r+1)  + o(s^0) \, ,  \nonumber \\
\nonumber \\
Q_{r,2}(s)&=&\frac {1}{8}(-1)^{\frac {r}{2}+1}L (r+2)(r+1)r\zeta (r+3)+r(-1)^{\frac {r}{2}+1}(L-3)\zeta (2) \zeta (r+1)+ \nonumber \\
&+& 2 (-1)^{1-\frac {r}{2}}\zeta (r)\{  2 \zeta (2)[ \ln s + \gamma _E-(L-2)\ln 2 ]+ 7\zeta (3)-2L \zeta (3)  \} - \nonumber \\
&-&  r(r+1) (-1)^{\frac {r}{2}}[ \ln s + \gamma _E-(L-2)\ln 2 ] \zeta (r+2) + \nonumber \\
&+&\frac {1}{2}(-1)^{\frac {r}{2}+1}(r+1)(r+2)L \zeta (r+3)+\nonumber \\
&+& 2 (-1)^{\frac {r}{2}+1}(r+1) [ \ln s + \gamma _E-(L-2)\ln 2 ] \zeta (r+2) + \nonumber \\
&+&\frac {1}{4} (-1)^{\frac {3r}{2}} (r+2) [-L \zeta
(r+3)+(L-2)(2^{r+3}-1)\zeta (r+3) ]  + o(s^0) \, . \nonumber
\end{eqnarray}
As far as we know these expansions are new results.

\subsection{The NLIE in the $s\rightarrow +\infty$ limit}

As in the one loop case, in the limit $s\rightarrow +\infty$ the
NLIE satisfied by the counting function reduces to the linear
equation satisfied by the density of roots, i.e. the BES equation.

The counting function satisfies the NLIE
\begin{eqnarray}
Z(u)&=&-2L \arctan 2u -i L \ln \left ( \frac {1+\frac
{g^2}{2{x^-(u)}^2}}{1+\frac {g^2}{2{x^+(u)}^2}} \right ) +
2\int _{-b}^{b} \frac {dv}{2\pi} \arctan (u-v)\frac {d}{dv} [Z(v) -2 L(v)] - \nonumber \\
&-& 2i \int _{-b}^{b} \frac {dv}{2\pi}  \left [ \ln \left ( \frac {1-\frac {g^2}{2x^+(u)x^-(v)} }{1-\frac {g^2}{2x^-(u)x^+(v)}} \right ) +i\theta (u,v) \right ]\frac {d}{dv} [Z(v) -2 L(v)] + \nonumber \\
&+& 2 \sum _{h=1}^{L-2} \arctan (u-u_h^{(i)}) -2i \sum _{h=1}^{L-2}
\left [ \ln \left ( \frac {1-\frac
{g^2}{2x^+(u)x^-(u_h^{(i)})}}{1-\frac {g^2}{2x^-(u)x^+(u_h^{(i)})}}
\right ) +i\theta (u,u_h^{(i)}) \right ] \, , \label {NLIE}
\end{eqnarray}
where $u_h^{(i)}$ refers to the holes present inside the interval
$[-b,b]$.

As in the one loop case, we go to the limit $s\rightarrow +\infty$.
In this limit, at the leading order in $s$ (i.e. $\ln s$) we can
drop the second and the last term in the rhs of the NLIE, as well as
also the terms containing $L(v)={\mbox {Im}}\ln [1+e^{iZ(v-i0)}]$.
We end up with the linear equation
\begin{eqnarray}
Z(u)&=&-2L \arctan 2u + 2\int _{-b}^{b} \frac {dv}{2\pi} \arctan (u-v)\frac {d}{dv}  Z(v)  - \label {Zlin}\\
&-& 2i \int _{-b}^{b} \frac {dv}{2\pi}  \left [ \ln \left ( \frac
{1-\frac {g^2}{2x^+(u)x^-(v)} }{1-\frac {g^2}{2x^-(u)x^+(v)}} \right
) +i\theta (u,v)\right ] \frac {d}{dv} Z(v)  + 2 \sum _{h=1}^{L-2}
\arctan (u-u_h^{(i)})  \, .  \nonumber
\end{eqnarray}
It seems natural to split the solution $Z(u)$ as
\begin{equation}
Z(u)=Z_0(u)+Z^{H}(u) \, ,
\end{equation}
where $Z_0(u)$ is the solution of the one loop part of (\ref
{Zlin}), already written in (\ref {Z0eq}) and $Z^{H}(u)$ is the
solution of the higher than one loop part of (\ref {Zlin}),
\begin{eqnarray}
Z^{H}(u)&=&-2i \int _{-b_0}^{b_0} \frac {dv}{2\pi} \left [ \ln \left ( \frac {1-\frac {g^2}{2x^+(u)x^-(v)} }{1-\frac {g^2}{2x^-(u)x^+(v)}} \right ) +i\theta (u,v) \right ]\frac {d}{dv}  Z_0(v) +  \label {many} \\
&+&2\int _{-\infty}^{+\infty} \frac {dv}{2\pi} \left [ \arctan
(u-v)-i \ln \left ( \frac {1-\frac {g^2}{2x^+(u)x^-(v)} }{1-\frac
{g^2}{2x^-(u)x^+(v)}} \right ) +\theta (u,v) \right ]\frac {d}{dv}
Z^{H}(v)  \, . \nonumber
\end{eqnarray}
Let us now define the densities
\begin{equation}
\rho _0(u)=-\frac {1}{2\pi s}\frac {d}{du}  Z_0(u)=\frac {1}{s} \bar
\rho _0 (\bar u) \, , \quad \sigma ^{H}(u)=-\frac {1}{2\pi s g^2}
\frac {d}{du} Z ^{H} (u) \, ,
\end{equation}
In terms of them, (\ref {many}) reads as follows
\begin{eqnarray}
0&=&2\pi \sigma ^{H}(u)- 2 \int _{-\infty}^{+\infty }dv \Bigl [ \frac {1}{1+(u-v)^2} -i \frac {d}{du} \ln \left ( \frac {1-\frac {g^2}{2x^+(u)x^-(v)} }{1-\frac {g^2}{2x^-(u)x^+(v)}} \right )+ \\
&+& \frac {d}{du}\theta (u,v) \Bigr ] \sigma ^{H}(v)+ \frac
{2i}{g^2} \int _{-b_0}^{b_0} dv \frac {d}{du} \left [ \ln \left (
\frac {1-\frac {g^2}{2x^+(u)x^-(v)} }{1-\frac {g^2}{2x^-(u)x^+(v)}}
\right )+i\theta (u,v)\right ] \rho _0 (v) \, . \nonumber
\end{eqnarray}
This equation coincides with the BES equation. In particular, if we
drop the dressing factor, it reduces to the ES equation - see (65)
of \cite {ES}.

\subsection{An alternative derivation of the anomalous dimension}

In the large spin limit, for the leading $O(\ln s)$ contribution,
$f(g)$, an elegant proportionality holds between the all-loops
energy (anomalous dimension) and the Fourier zero mode of the
higher-than-one-loop density of roots \cite{KL}. Here, we are going
to show that the connection extends up to the $O(s^0)$ order.

Regardless the normalisation, we may define the higher than one
loop and one loop densities as $\sigma ^H(u)=\frac {d}{du}F^H(u)$,
$\sigma _0(u)=\frac {d}{du}F_0(u)$, respectively, and write down
their specific linear integral equations in the Fourier space. For we
may move from (\ref{Ftildes}) and use the fundamental Fourier
transforms
\begin{equation}
\int _{-\infty}^{\infty} du e^{-iku} \left [\frac
{1}{x^{\pm}(u)}\right ]^r= \pm r \left (\frac {\sqrt {2}}{ig} \right
)^r \theta (\pm k)\frac {2\pi}{k} e^{\mp \frac {k}{2}}J_r({\sqrt
{2}}gk) \, , \label {xpmfour}
\end{equation}
to obtain
\begin{eqnarray}
\hat \sigma ^H(k)&=&\pi L \frac {1-J_0({\sqrt {2}}gk)}{\sinh \frac
{|k|}{2}}+
\nonumber \\
&+& \frac {1}{2 \sinh \frac {|k|}{2}}\int _{-\infty}^{\infty} dh
\Bigl [ \sum _{r=1}^{\infty} \frac {r}{|h|}(-1)^{r+1}J_r({\sqrt
{2}}gk)
J_r({\sqrt {2}}gh)(1-{\mbox {sgn}}(kh))e^{-\frac {|h|}{2}}+ \nonumber \\
&+& 2\sum _{r=2}^{\infty}\sum _{\nu
=0}^{\infty}c_{r,r+1+2\nu}(g)(-1)^{r+\nu} \frac {e^{-\frac
{|h|}{2}}}{h} \Bigl (J_{r-1} ({\sqrt {2}}gk)J_{r+2\nu}
({\sqrt {2}}gh)- \nonumber \\
&-& J_{r-1}({\sqrt {2}}gh)J_{r+2\nu}({\sqrt {2}}gk)\Bigr ) \Bigr ]
[\hat \sigma ^H(h)+\hat \sigma _0(h)+2\pi (L-2)] +o(s^0) \, .
\nonumber
\end{eqnarray}
Performing now the limit $k\rightarrow 0^{\pm}$, we gain
\begin{equation}
\lim _{k\rightarrow 0^+}\hat \sigma ^H(k)= \lim _{k\rightarrow
0^-}\hat \sigma ^H(k)=-\frac {g}{\sqrt {2}} \int
_{-\infty}^{\infty}\frac {dh}{h} J_1({\sqrt {2}}gh)e^{-\frac
{|h|}{2}} [\hat \sigma ^H(h)+\hat \sigma _0(h)+2\pi (L-2)] +o(s^0)
\, . \label {limsigma}
\end{equation}
On the other hand, if we re-write the energy (\ref{egs}) up to the
order $s^0$ by means of the Fourier transforms (\ref{xpmfour}), we
also obtain
\begin{equation}
E(g,s)=-\frac {1}{\sqrt {2}\pi g } \int _{-\infty}^{\infty}\frac
{dh}{h} J_1({\sqrt {2}}gh)e^{-\frac {|h|}{2}} [\hat \sigma
^H(h)+\hat \sigma _0(h)+2\pi (L-2)] + o(s^0) \, . \label {egs1}
\end{equation}
Upon comparing (\ref {limsigma}) and (\ref {egs1}), we size the
desired relation
\begin{equation}
\hat \sigma ^H(0)=\pi g^2 E(g,s)+o(s^0) \, . \label {e-sigma}
\end{equation}

\medskip

Now, in order to check this result we can repeat the perturbative
expansion (\ref{bf}) and extract from it the quantities $E_0(s)$ and
$E_1(s)$. In fact, (\ref{FH1}) yields, at order $g^2$, the Fourier
transform
\begin{equation}
\hat \sigma _1^{H}(k)=\frac{1}{2 \sinh(\frac{|k|}{2})}\left[\pi L
k^2+ 4\pi |k|(\ln s + \gamma _E-(L-2)\ln 2)\right] \, ,
\end{equation}
which just verifies
\begin{equation}
\hat \sigma ^H_1(0)=\pi E_0(s) \, . \label {esigma1}
\end{equation}
Similarly, (\ref{FH2}) implies
\begin{eqnarray}
&&\hat \sigma _2^{H}(k)=-\frac{1}{2 \sinh(\frac{|k|}{2})}\Bigl \{
\frac{\pi L  k^4}{8}+\frac{ \pi |k|^3E_0(s)}{4}+\frac{ \pi^3 (L-3)k^2}{6}+\frac{\pi^3 |k|E_0(s)}{6} + \nonumber\\
&+& 2 \pi |k|[7\zeta(3)-2L\zeta(3)]\Bigr \} \, .
\end{eqnarray}
Finally, for $k=0$
\begin{equation}
\hat \sigma _2^{H}(0)=-\frac{\pi^3E_0(s)}{6}-2
\pi[7\zeta(3)-2L\zeta(3)]= \pi E_1(s) \, .   \label {esigma2}
\end{equation}
Of course, these results, (\ref {esigma1}, \ref {esigma2}), agree
with the general one (\ref {e-sigma}).

\section{Summary}

In this paper we have developed and applied the technique of the
NLIE on intervals sketched in \cite{BFR}. This new formalism allows
to treat magnon scattering matrices with general dependence on the
rapidities and states with roots on intervals of the real line (or
even of complex lines). Therefore, it seems perhaps more indicated than the
historical method presented in \cite {FMQR}, if we want to study Bethe
equations appearing in the context of ${\cal N}=4$ SYM.

We have given an explicit application of the NLIE on interval for the Bethe
Ansatz type equations describing the $sl(2)$ sector of ${\cal N}=4$
SYM. In particular, we have written the exact equations which allow us to define the relevant
functions (i.e. the forcing term $F$ and the kernel function $G$)
entering the NLIE. Then, we passed on to studying the limit of large number
of Bethe roots (or spin $s$). In this limit and at the
leading order $\ln s$, the NLIE as well as the equation for the
derivative of the forcing term naturally becomes the BES
equation. If we take into account also the sub-leading correction
$O(s^0)$, the forcing term satisfies a modification of the BES
equation, with a different inhomogeneous term. Therefore, this equation is
suitable also for a (non-perturbative and) strong coupling study, even
if possible results in this direction should be corrected by
eliminating the wrapping effects. Interestingly, we noticed that in
the formalism of the NLIE on intervals the non-linear terms are
negligible as $s\rightarrow +\infty$. Therefore, in order to
determine the eigenvalues of the conserved charges up to the order
$s^0$, it is sufficient to consider in their general expressions only
linear terms, involving the forcing term. This conceptually and
practically enormously simplifies their calculations and could
suggest applications of our formalism to cases in which wrapping
effects are under control or absent.

\medskip

{\bf Acknowledgements} We acknowledge the INFN grant "Iniziativa
specifica PI14" {\it Topics in non-perturbative gauge dynamics in
field and string theory} for travel financial support. M.R. thanks
the INFN-Bologna and the Department of Physics for generous
hospitality.

\appendix
\section{Numerical evaluation of the non-linear integrals}
\setcounter{equation}{0} In this appendix, we will analyse the large
spin $s$ behaviour of the non-linear integral function in
(\ref{2eq1}) and of the non-linear integral in (\ref{Oexp}). Both
involve the {\it logarithmic} function (of $Z(u)$), $L(u)$, as
defined in (\ref{L}). In principle, we could pursue an analytic
saddle point evaluation as in the sine-Gordon case \cite{FMQR},
since anew, because of its forcing term, the counting function $Z$
scales like the {\it size} $\ln s$, thus implying , roughly
speaking, a correction of order $\sim e^{-\ln s}=1/s$: we will see
that this easy conclusion should be not far from reality.
Nevertheless, we find here more instructive for us to gain some
flavour by performing a still easy numerical exercise. In fact, for
simplicity's sake we restrict ourselves to the one loop case and, in
the aforementioned spirit, we approximate $Z_0(u)$ in $L_0(u)$ by
the forcing term $F_0(u)$ as given by (\ref{Fkappa}). In the
numerical implementation, we must learn to write the integrals for
finite $\epsilon$: thanks to the Cauchy theorem, we may move the
integration contour in the complex plane as far as we do not meet a
singularity
\begin{eqnarray}
\delta Z_0(u;s)&=&-\mbox{Im}\int _{-b_0(s)}^{b_0(s)}\frac{dv}{\pi}\,\phi_0(u,v-i\epsilon)\,\frac {d}{dv}\ln\left[1+e^{iZ_0(u-i\epsilon)}\right]-\label{counting}\\
&-&\mbox{Im}\int _{0}^{\epsilon}\frac {dy}{\pi}\left\{ \phi_0(u,-b_0(s)-iy)\,\frac {d}{dy}\ln\left[1+e^{iZ_0(-b_0(s)-iy)}\right]-\right.\nonumber\\
&-&\left.\phi_0(u,b_0(s)-iy)\,\frac
{d}{dy}\ln\left[1+e^{iZ_0(b_0(s)-iy)}\right]\right\} \, ,\nonumber
\end{eqnarray}
with $\phi_0(u,v)$ defined in (\ref{sl2funct}) and $b_0(s)$
determined by the condition (\ref{Zcond}), which gives
$b_0(s)\simeq0.472\,s$. In fact, if we should assume exactly
$b_0(s)=s/2$, we would observe an external hole jumping into the
interval of integration at a certain value of $s$. Of course, this
would give a deceiving discontinuous dependence of the integrals
(\ref{counting}) on $s$. In other words, the value $s/2$ is not a
good one as for the separator \cite{FR} between the roots and the
external hole. Once clarified this crucial point, we may evaluate
the integrals with \texttt{Mathematica}. For instance, we report in
Fig. 1 the behaviour of (\ref{counting}) with $s$ on the $x$-axis,
for $u=1$ and $\epsilon=0.1$: the leading $\sim 1/s$ decrease can be
easily spotted.
\begin{figure}[h]
\begin{center}
\includegraphics[height=0.5\textwidth,width=0.9\textwidth]{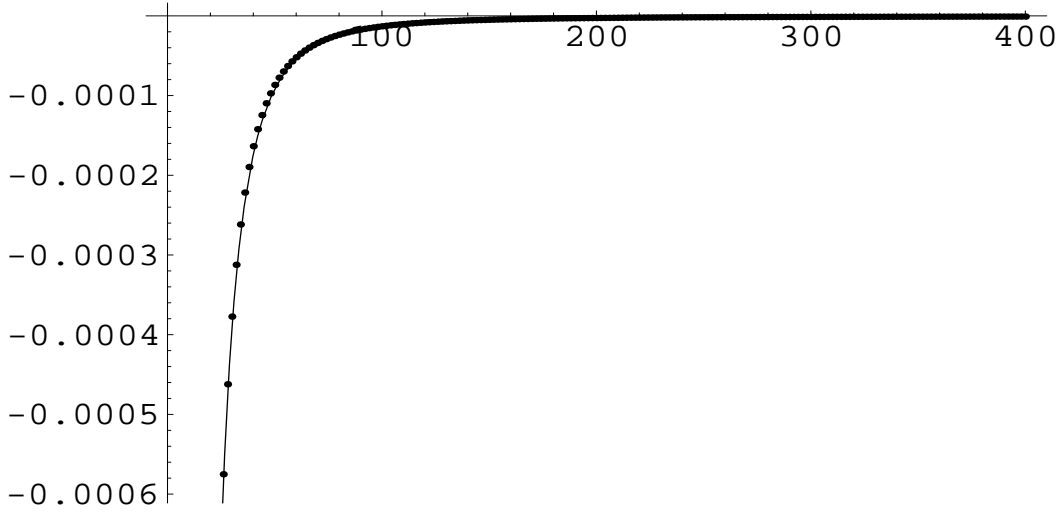}
\caption{Comparison between numerical evaluations (dots) of $\delta
Z_0(u=1;s)$ and their best fit as $a_1/s+a_2/s^2+a_3/s^3$ (line), with
$a_1\simeq0.001$, $a_2\simeq-0.105$, $a_3\simeq-8.137$.}
\end{center}
\end{figure}
Eventually, we want to numerically evaluate the non-linear integral
in the one-loop energy, namely the last line of (\ref{Oexp}), once
we specialise the observable $O(v)=q_2(v)$. We still need to take
into account the "lateral" contributions due to the integration over
($-i\,\epsilon$,$i\,\epsilon$) and then, upon using the leading
order solution $G_0(v,w)$ of Section 3.3, we obtain
\begin{eqnarray}
\delta E_0(s)&=&-\mbox {Im}\int _{-b_0(s)}^{b_0(s)}\frac
{dv}{\pi}\,q_2(v)\,\int _{-b_0(s)}^{b_0(s)} dw
\frac{d}{dv}\,\left[G_0(v,w-i\epsilon)\right.-\\
&-&\left.\delta(v-w+i\,\epsilon)\right]{}\ln \left [1+e^{iZ_0(w-i\epsilon)}\right]+\nonumber\\
&+&\int _{0}^{\epsilon}\frac {dy}{\pi}\left\{ \frac{d}{dv}\left[G_0(v,-b_0(s)-iy)-\right.\right.\nonumber\\
&-&\left.\left.\delta(v+b_0(s)+iy)\right]\frac {d}{dy}\ln \left
[1+e^{iZ_0(-b_0(s)-iy)}\right]\right.\nonumber\\
&-&\left.\frac{d}{dv}\left[G_0(v,b_0(s)-iy)-\delta(v-b_0(s)+iy)\right]\frac
{d}{dy}\ln \left [1+e^{iZ_0(b_0(s)-iy)}\right]\right\} \, .
\nonumber\label{int2}
\end{eqnarray}
Actually, since the counting function is odd, we may substitute the
$v$-derivative of $G_0(v,w)$ directly the antisymmetric combination
(\ref{X}). The dependence of $\delta E_0(s)$ on $s$ is plotted in
Fig. 2, where we still have $\epsilon=0.1$, and also in this case
the behaviour seems to be suitably fitted by a polynomial of $1/s$,
yielding again a $1/s$ leading contribution. Unfortunately, this
one loop set-up cannot exhibit the appearance of the logarithms 
$\ln s$, in agreement with the large $s$ expansions of \cite{Bec, KLRSV, BDM}.
\begin{figure}[h]
\begin{center}
\includegraphics[height=0.5\textwidth,width=0.9\textwidth]{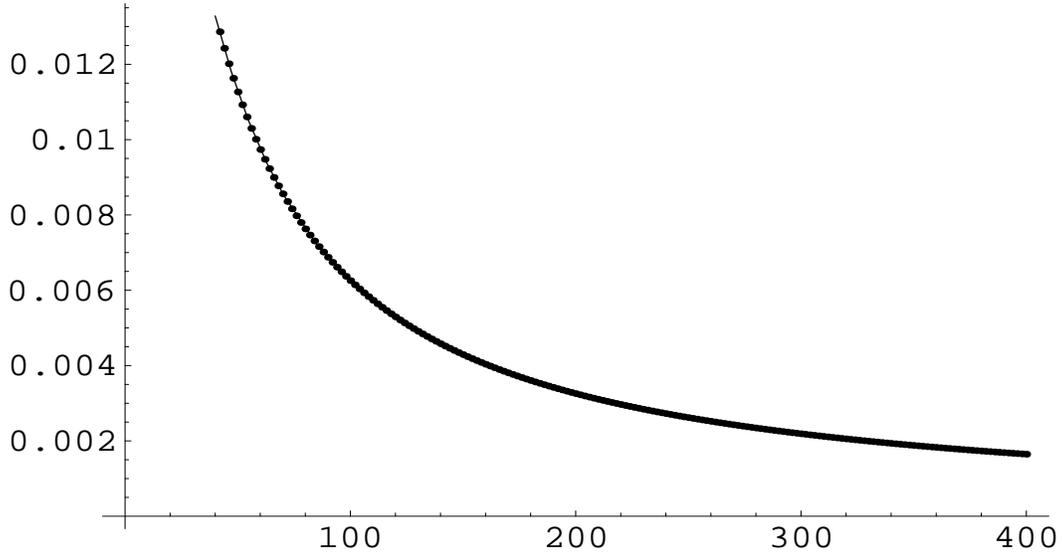}
\caption{Comparison between numerical evaluations (dots) of $\delta
E_0(s)$ and their best fit as $b_1/s+b_2/s^2+b_3/s^3$ (line), with
$b_1\simeq0.675$, $b_2\simeq-4.667$, $b_3\simeq-39.390$.}
\end{center}
\end{figure}

\end{document}